\documentstyle{merrittapj}

\newif\ifAMStwofonts

\def\fun#1#2{\lower3.6pt\vbox{\baselineskip0pt\lineskip.9pt
  \ialign{$\mathsurround=0pt#1\hfil##\hfil$\crcr#2\crcr\sim\crcr}}}
\def\lap{\mathrel{\mathpalette\fun <}}
\def\gap{\mathrel{\mathpalette\fun >}}
\def\mass{{\cal M}}
\def\Msolar{{\mass_\odot}}

\def\Lsolar{{L_\odot}}

\def\h2{$\rm H_2\,$}

\def\n2h{ H$\alpha$+[N~II]}

\def\beq{\begin{equation}}
\def\eeq{\end{equation}}

\def\bc{\begin{center}}
\def\ec{\end{center}}

\ifoldfss
  \ifCUPmtlplainloaded \else
    \NewTextAlphabet{textbfit} {cmbxti10} {}
    \NewTextAlphabet{textbfss} {cmssbx10} {}
    \NewMathAlphabet{mathbfit} {cmbxti10} {} % for math mode
    \NewMathAlphabet{mathbfss} {cmssbx10} {} %  "   "    "
  \fi
  \ifAMStwofonts
    \ifCUPmtlplainloaded \else
      \NewSymbolFont{upmath} {eurm10}
      \NewSymbolFont{AMSa} {msam10}
      \NewMathSymbol{\upi}     {0}{upmath}{19}
      \NewMathSymbol{\umu}     {0}{upmath}{16}
      \NewMathSymbol{\upartial}{0}{upmath}{40}
      \NewMathSymbol{\leqslant}{3}{AMSa}{36}
      \NewMathSymbol{\geqslant}{3}{AMSa}{3E}

      \let\leq=\leqslant \let\le=\leqslant
      \let\geq=\geqslant \let\ge=\geqslant
    \fi
  \fi
\fi % End of OFSS

\ifnfssone
  \newmathalphabet{\mathit}
  \addtoversion{normal}{\mathit}{cmr}{m}{it}
  \addtoversion{bold}{\mathit}{cmr}{bx}{it}
  \newmathalphabet{\mathbfit} % math mode version of \textbfit{..}
  \addtoversion{normal}{\mathbfit}{cmr}{bx}{it}
  \addtoversion{bold}{\mathbfit}{cmr}{bx}{it}
  \newmathalphabet{\mathbfss} % math mode version of \textbfss{..}
  \addtoversion{normal}{\mathbfss}{cmss}{bx}{n}
  \addtoversion{bold}{\mathbfss}{cmss}{bx}{n}
  \ifAMStwofonts
    \ifCUPmtlplainloaded \else
      \UseAMStwoboldmath
      \makeatletter
      \new@mathgroup\upmath@group
      \define@mathgroup\mv@normal\upmath@group{eur}{m}{n}
      \define@mathgroup\mv@bold\upmath@group{eur}{b}{n}
      \edef\UPM{\hexnumber\upmath@group}
      \new@mathgroup\amsa@group
      \define@mathgroup\mv@normal\amsa@group{msa}{m}{n}
      \define@mathgroup\mv@bold\amsa@group{msa}{m}{n}
      \edef\AMSa{\hexnumber\amsa@group}
      \makeatother
      \mathchardef\upi="0\UPM19
      \mathchardef\umu="0\UPM16
      \mathchardef\upartial="0\UPM40
      \mathchardef\leqslant="3\AMSa36
      \mathchardef\geqslant="3\AMSa3E

      \let\leq=\leqslant \let\le=\leqslant
      \let\geq=\geqslant \let\ge=\geqslant
    \fi
  \fi
\fi % End of NFSS release 1

\ifnfsstwo
  \DeclareMathAlphabet{\mathbfit}{OT1}{cmr}{bx}{it}
  \SetMathAlphabet\mathbfit{bold}{OT1}{cmr}{bx}{it}
  \DeclareMathAlphabet{\mathbfss}{OT1}{cmss}{bx}{n}
  \SetMathAlphabet\mathbfss{bold}{OT1}{cmss}{bx}{n}
  \ifAMStwofonts
    \ifCUPmtlplainloaded \else
      \DeclareSymbolFont{UPM}{U}{eur}{m}{n}
      \SetSymbolFont{UPM}{bold}{U}{eur}{b}{n}
      \DeclareSymbolFont{AMSa}{U}{msa}{m}{n}
      \DeclareMathSymbol{\upi}{0}{UPM}{"19}
      \DeclareMathSymbol{\umu}{0}{UPM}{"16}
      \DeclareMathSymbol{\upartial}{0}{UPM}{"40}
      \DeclareMathSymbol{\leqslant}{3}{AMSa}{"36}
      \DeclareMathSymbol{\geqslant}{3}{AMSa}{"3E}

      \let\leq=\leqslant \let\le=\leqslant
      \let\geq=\geqslant \let\ge=\geqslant
    \fi
  \fi
\fi % End of NFSS release 2

\ifCUPmtlplainloaded \else
  \ifAMStwofonts \else % If no AMS fonts
    \def\upi{\pi}
    \def\umu{\mu}
    \def\upartial{\partial}
  \fi
\fi

\title{Regular and Chaotic Dynamics of Triaxial Stellar Systems}
\author[Monica Valluri and David Merritt]
       {Monica Valluri and David Merritt\\
Department of Physics and Astronomy, Rutgers University,
New Brunswick, NJ 08855.}
\date{Rutgers Astrophysics Preprint Series No. 214}

\pubyear{1997}

\begin{document}

\maketitle

\label{firstpage}

\begin{abstract}
We use Laskar's frequency mapping technique to study the dynamics of
triaxial galaxies with central density cusps and nuclear black holes.  
For ensembles of $\sim 10^4$ orbits, we
numerically compute the three fundamental frequencies of the motion,
allowing us to map out the Arnold web.  
We also compute diffusion rates of stochastic orbits in frequency space.
The objects of greatest importance in structuring phase space are found 
to be the 3-dimensional resonant tori, regions where 
the fundamental frequencies satisfy a relation of the form
$0=l\omega_1+m\omega_2+n\omega_3$ with integer $(l,m,n)$.  
When stable, resonant tori generate phase space regions 
in which the motion is regular; these regions are not 
necessarily associated with a stable periodic orbit
as in systems with only two degrees of freedom.  
Boxlike orbits are generically stochastic but some tube
orbits are stochastic as well.  The spectrum of diffusion rates for
boxlike orbits at a given energy is well approximated as a power law
over at least six decades.
Models with high central
concentrations -- steep central cusps or massive black holes --
exhibit the most stochasticity.  
Even a modest black hole, with a mass
$\sim 0.3\%$ the mass of the galaxy, is as effective as the
steepest central density cusp at inducing stochastic diffusion.  There
is a transition to global stochasticity in boxlike phase space when
the mass of a central black hole exceeds $\sim 2\%$ of the galaxy mass.
We estimate the dependence of orbital evolution rates on galaxy
structural parameters.
We predict a greater average degree of dynamical evolution in faint 
elliptical galaxies due to their high central densities and short crossing
times.  
The evolution time is estimated to be shorter than a galaxy lifetime 
for absolute magnitudes fainter than about $-19$ or $-20$, consistent 
with the observed change in many elliptical galaxy properties at this 
luminosity.
\end{abstract}

\section{Introduction}

Motion in a smooth gravitational field becomes quite simple if the
number of isolating integrals equals or exceeds the number of degrees
of freedom, and much work in galactic dynamics has focussed on finding
integrable or near-integrable models for galactic potentials.  
Kuzmin (1956, 1973) showed that there is a unique,
ellipsoidally-stratified mass model for which the corresponding
potential has three global integrals of the motion, quadratic in the
velocities.  Kuzmin's model -- explored in detail by de Zeeuw (1985)
who christened it the ``Perfect Ellipsoid'' -- has a large,
constant-density core in which the orbital motion is that of a 3-D
harmonic oscillator.  Every orbit in the core of the Perfect Ellipsoid
fills a rectangular parallelepiped, or box.  At higher energies in the
Perfect Ellipsoid, box orbits persist and three new orbit families
appear: the tubes, orbits that preserve the direction of their
circulation around either the long or short axis of the figure.  Tube
orbits respect an integral of the motion analogous to the angular
momentum, and hence -- unlike box orbits -- avoid the center.  
All integrable triaxial potentials have a similar orbital 
structure (\cite{hun95}). 

The large core of the Perfect Ellipsoid is a poor
match to real elliptical galaxies, all of which exhibit power-law
cusps in the stellar density at small radii (\cite{fer94}; 
\cite{mef95}; \cite{geb96}).  
There is increasingly strong evidence that many elliptical galaxies and bulges
also contain massive central objects, possibly the black holes that
are thought to have powered quasars (\cite{kor95}).  While the masses
of these dark central components are often very uncertain, typical
estimates are $10^{-3}\lap M_h/M_g\lap 10^{-2}$, where $M_h$ is the
black hole mass inferred from the orbital motions of surrounding stars
and gas and $M_g$ is the stellar mass of the host galaxy or (in the
case of a spiral galaxy) the mass of the stellar bulge.  Some
galaxies, like M32, the dwarf companion to the nearby Andromeda
galaxy, are known to contain both a steep stellar cusp ($\rho\propto
r^{-1.6})$ and a dynamically-significant black hole ($M_h/M_g\sim
0.003)$.

The purpose of the present study is to explore the orbital 
dynamics of realistic triaxial potentials in a systematic way.
One goal is to understand how the structure of triaxial phase 
space differs from that of axisymmetric systems.
Another is to estimate the rate at which chaos 
implies changes in the structure of a triaxial galaxy, and to 
understand how this rate depends on galaxy properties.
In a pioneering study, Goodman \& Schwarzschild (1981) found that
the boxlike orbits in a triaxial model with a smooth core were
often stochastic, but that the behavior of these orbits was essentially
regular for at least $50$ oscillations.
On the other hand, Merritt \& Fridman (1996) found that the
stochasticity in triaxial models with $\rho\propto r^{-2}$ density 
cusps produced significant changes in the appearance of boxlike 
orbits after just a few tens of oscillations.
Merritt \& Valluri (1996) went on to calculate timescales for
mixing in these strongly chaotic potentials; they found that
ensembles of stochastic trajectories could evolve toward invariant
distributions -- corresponding to an approximately uniform 
filling of stochastic phase space -- on timescales of only $10^1-10^2$
orbital periods.
Kandrup and coworkers (\cite{kam94}; \cite{mah95}) noted similar, 
rapid rates of chaotic mixing in a variety of two-dimensional systems.
Taken together, these results suggest that the characteristic 
time over which chaos manifests itself in the orbital motion of a 
triaxial galaxy is strongly dependent on its radial mass distribution.
Presumably, this dependence reflects the sensitivity of boxlike 
orbits to deflections that occur during close passages to the galaxy 
center.

Here we use Laskar's frequency mapping technique (\S2) to look in much 
more detail at the structure of regular and chaotic phase space 
in realistic triaxial potentials.
Laskar's technique -- summarized in more detail below -- allows 
one to quickly and accurately compute the fundamental frequencies 
that characterize the quasi-periodic motion of regular orbits on 
their invariant tori.
In addition, one can compute the rate at which these frequencies 
change as a stochastic orbit diffuses from one torus to another.
The entire phase space at a given energy can then be represented 
on a single diagram, the frequency map, which shows the way in 
which an otherwise-integrable phase space is modified by the 
existence of resonances between the three degrees of freedom.
The resonances that are most important for determining the 
structure of phase space can immediately be picked out.

The models considered in this study are the triaxial 
generalizations of the spherical models discussed 
by Dehnen (1993), Carollo (1993) and Tremaine {\it et al.} (1994).
These models have a mass density
\beq
\rho(m) = {(3-\gamma) M\over 4\pi abc} m^{-\gamma} (1+m)^{-(4-\gamma)},
\ \ \ \ 0\le\gamma < 3 
\eeq
with
\beq
m^2={x^2\over a^2} + {y^2\over b^2} + {z^2\over c^2}, \ \ \ \ a\ge b\ge c\ge
0, 
\eeq
and $M$ the total mass.
The mass is stratified on ellipsoids with axis ratios $a:b:c$; the $x$ [$z$]
axis is the long [short] axis.
The parameter $\gamma$ determines the slope of the central density cusp.
For $\gamma=0$ the model has a finite-density core, while for
all $\gamma>0$ the central density is infinite.
The strongest cusps that we will consider here have $\gamma=2$, i.e.
$\rho\propto m^{-2}$ at small radii. 
Henceforth we will refer to this model as a ``$\gamma$-model'' or 
as ``Dehnen's model.''
In the spherical case, the radial force in this model is
\beq
-{\partial\Phi\over\partial r}=-{GM\over a^2}\left({r\over a}\right)^
{1-\gamma}\left(1+{r\over a}\right)^{\gamma-3},
\eeq
which is finite at the center for $\gamma<1$ (``weak cusp'') and
divergent for $\gamma\geq 1$ (``strong cusp'').  
To represent the effect on the stellar motions of a central black
hole, we also consider models with an added central point of mass
$M_h$. 
The $\gamma$-model is a reasonable description of the stellar density
at small and intermediate radii in elliptical galaxies.
Our choice of this model is motivated by the expectation
that the central regions of a triaxial galaxy are the most 
important for determining the behavior of the boxlike
orbits.
One of our principal objectives will be to establish whether there is a 
critical value of cusp slope $\gamma$ or black hole mass $M_h$ at 
which a globally significant transition occurs in the structure of
phase space.

We find $(\S 3)$ that the motion of boxlike orbits in triaxial
$\gamma$-models
is generically chaotic, but that the timescale for diffusion in
stochastic phase space exhibits great variation, both within a single
model and as a function of $\gamma$ and $M_h$.  We confirm the result
of Papaphilippou \& Laskar (1998) that the objects of fundamental
importance for structuring phase space are the 3-dimensional resonant
tori; in this sense, 3DOF systems differ from 2DOF systems, in which
periodic orbits and their associated families give phase space its
structure.  As the central concentration of the models is increased,
boxlike phase space becomes more and more chaotic.  
A transition to global stochasticity occurs ($\S4$) when the 
central mass contains between $\sim 2\%$ of the galaxy mass; 
for such large values of
$M_h$, the boxlike phase space is almost completely stochastic, and
diffusion takes place in a very short time.  
Interestingly, the critical black hole mass that we find for 
transition to global stochasticity is close to the maximum mass 
observed in real galaxies (\cite{kor95}),
and to the mass that induces a sudden evolution toward
axisymmetry in $N$-body simulations (\cite{meq98}).
Central density cusps without black holes are less effective at 
generating chaos; even a modest black hole, with $M_h\approx0.3\%M_g$, 
induces about as much stochastic diffusion as the steepest 
($\rho\propto r^{-2})$ cusps.

Following Laskar (1993), we characterize the diffusion in phase space 
via the rate of change of the characteristic frequencies computed over
a fixed interval of time.
We find that the spectrum of diffusion rates at a given 
energy can be well approximated as a power law with slope close 
to $-1$; this spectrum extends over at least six decades ($\S5$).
Thus, every triaxial potential examined here has a significant
population of slowly-evolving stochastic orbits.  After a long enough
time, even weakly stochastic orbits would be expected to contribute to
chaotic mixing, an essentially irreversible process that leads to
steady-state, invariant distributions in stochastic phase space. 
Once mixing sets in,
triaxiality is likely to disappear, since regular box orbits are
necessary for maintaining triaxial shapes.  We estimate ($\S6$) the
timescale for this to take place as a function of galaxy luminosity;
because typical orbital periods are shorter in faint ellipticals, they
should be dynamically ``more evolved'' than bright ellipticals and
hence less likely to maintain triaxial shapes.  We predict evolution
times that are roughly equal to galaxy lifetimes for elliptical
galaxies with $M_B\approx -19$ to $-20$, close to the absolute magnitude at
which many elliptical galaxy properties -- including the 
distribution of intrinsic shapes -- are observed to change.  
We are therefore led to a picture ($\S7$) in which many of the systematic 
differences between bright and faint ellipticals 
are a consequence of their different dynamical ages.

\section{Numerical Techniques}

Laskar (1990) developed a new technique, the numerical analysis of
fundamental frequencies (NAFF), for analyzing chaotic motion in
the solar system.  Laskar's technique is based on the principle that
an orbit in a system for which the motion is close to integrable is
described by a well-defined set of frequencies $\omega_i$, one per
degree of freedom.  For a regular orbit, the fundamental frequencies
are invariant with respect to the coordinates in which the motion is
represented.  
An analysis of the variation of the fundamental
frequencies with respect to time enables one to detect the
presence of chaos on shorter timescales than are usually possible with
more classical techniques like Liapunov exponents.  In addition, the
NAFF algorithm allows one to quantify the degree to which a stochastic
orbit has evolved over a given period of time; by contrast, Liapunov
exponents reveal only the time-averaged rate of divergence in the
vicinity of the orbit.
A plot of the fundamental frequencies of a set of orbits drawn from a
regular grid of initial conditions yields the ``frequency map,''
effectively a representation of the Arnold web of the system.  The
frequency map permits an identification of the resonant layers
associated with stable and unstable motion as well as a determination
of the size of the chaotic zones.

Laskar and collaborators have applied the NAFF algorithm to a wide
variety of systems of astronomical and non-astronomical interest
(\cite{las90}; \cite{las92}; \cite{las93}; \cite{pal96}, 1998).  
These authors have shown
that the frequency analysis technique is a remarkably powerful way to
probe the structure of phase space in complex, non-integrable systems.

In the following subsections, we give a brief overview of Hamiltonian
systems; summarize the basic principles of Laskar's technique; and
discuss several factors that affect the accuracy with which the
fundamental frequencies can be recovered.

The models that we consider, both here and below, are ``maximally
triaxial'' $\gamma$-models, i.e. they satisfy
\beq
{a^2-b^2\over a^2-c^2} = {1\over 2}.
\eeq
Henceforth we adopt units in which the total mass $M$, the $x$-axis
scale length $a$, and the gravitational constant $G$ are unity.  The
mass of the central black hole, when present, will be expressed as
$M_h$; since the total mass of the triaxial model is equal to one in
our units, $M_h$ may be interpreted as the mass of the black hole in
units of the galaxy mass.  Expressions for the gravitational potential
and forces are given by Merritt \& Fridman (1996).  Following those
authors, we assign orbits one of a set of 20 energies, defined as the
values of the potential on the $x$-axis of a set of ellipsoidal shells
- with the same axis ratios as the density - that divide the model
into 21 sections of equal mass.  Thus shell 1 encloses 1/21 of the
total mass, shell 2 encloses 2/21, etc.; shell 21 lies at infinity.
The period of the $x$-axis orbit will often be used as a reference
time; $\omega_0$ is defined as $2\pi/T_x$, the frequency of the axial
orbit.

\subsection{Frequencies in Hamiltonian Systems}

If a Hamiltonian system with {\it N} degrees of freedom (DOF) is 
integrable, the Hamiltonian $H({\bf J, \theta})$ can be written purely in
terms of the $N$ actions $J_j$, $H({\bf J, \theta}) =
H(J_j)$. 
The equations of motion of the system are then 
\beq
\dot{J_j} = 0, \qquad \; \; \dot\theta_j = {{\partial H}\over{\partial
J_j}} = \omega_j({\bf J}), 
\eeq 
for $j = 1,2,...,N$. 
Orbits in the system can be written in terms of the complex variables
\beq 
z_j(t) = J_je^{i\theta_j} = z_j(0) e^{i\omega_j t}.
\eeq 
If ${{\partial\omega_j({\bf J})}/{\partial J_J}} \neq 0$, the
system is non-degenerate and in principle the actions can be written
as 
\beq J_j = F_j(\omega_1,\omega_2,...,\omega_N).  
\eeq 
When a system is integrable, the motion of a particle projected onto the
$(J_j,\theta_j)$ plane describes pure circles.  
This is because the motion in phase space occurs on the surface of 
tori that are the products of true circles of radii $|J_j|$. 
The motion around a torus occurs at a rate determined by a frequency vector
$(\omega_1,\omega_2,...,\omega_N)$ which is fixed for each torus. 
In general, we do not know the action-angle variables $(J_j, \theta_j)$,
but in restricted cases close approximations $(J_j^\prime,
\theta_j^\prime)$ to these variables may be known.  
In these coordinates $(J_j^\prime, \theta_j^\prime)$ the motion
is no longer pure circles but it is still determined by the same 
vector of frequencies 
$\dot\theta_j^\prime = {{\partial H}/{\partial J_j^\prime}} =
\omega_j({\bf J})$ provided the new variables are canonically
conjugate.  
The new coordinates may be written as a Laurent
series expansion in terms of the action-angle variables $z_j(t)$:
\beq
\zeta(t) = z_j(t) + \sum_{\bf m} 
a_{\bf m}e^{i({\bf m\cdot\omega}) t}.
\label{zeta}
\eeq
In the limit that the coordinates $(J_j^\prime, \theta_j^\prime)$ are
action-angle variables, the amplitudes $a_{\bf m}$ tend to zero.

Realistic systems with more than one DOF are rarely
integrable, i.e. are rarely characterized by $N$ global invariants.
In some cases, the Hamiltonian can be written 
as a perturbation of an integrable Hamiltonian $H_0$, 
\beq
 H({\bf J,\theta}) = H_0({\bf J}) + \epsilon H_1({\bf J,\theta}). 
\eeq 
If the perturbation $\epsilon$ is small, the KAM 
theorem guarantees that a large fraction
of the tori still persist and the motion of most of the orbits is
still quasi-periodic. 
The motion on this discontinuous set of tori 
(defined by a Cantor set of frequency vectors \
${ \Omega} = (\omega_1,\omega_2,...,\omega_n) $ 
and therefore referred to as Cantori) 
is still describable in terms of action-angle variables.
However it is no longer possible to define a single set of action 
variables that are valid over the entire phase space as in 
equation (7), since the frequencies change discontinuously wherever 
the tori are destroyed.
Nevertheless, fundamental frequencies continue to exist for all
regular orbits.

A central role is played by the resonant tori, tori on which the
fundamental frequencies satisfy a relation like ${\bf m}\cdot\omega=0$
where ${\bf m}$ is a vector with integer components. 
The KAM theorem guarantees that ``very nonresonant'' tori -- tori 
for which ${\bf m}\cdot\omega$ is large -- 
persist under small perturbation of any integrable Hamiltonian.
Most of phase space is occupied by very nonresonant tori, but 
resonant tori are also dense in the phase space and do not 
survive small perturbations.
Stable resonant tori (or ``elliptic manifolds'') generate regions of 
regular motion while unstable resonant tori (``hyperbolic manifolds'') 
are often associated with stochastic layers.  
According to the Poincar\'e-Birkhoff theorem,
elliptic and hyperbolic resonances come in pairs, leading to
alternating regions of regular and stochastic motion.
While resonant tori are dense in the phase space, only those
of sufficiently low order -- i.e., those with $|{\bf m}|$ sufficiently
small -- are likely to affect more than a very small phase space
region.
Resonances appear as lines in the frequency map, and the strength of a 
particular resonance is immediately apparent from the degree to which 
the otherwise-regular grid of points has been distorted (\cite{las93}).
 
In a 2 DOF system, motion on resonant tori is closed, since the 
resonance condition implies $\omega_1/\omega_2 = -m/l$ guaranteeing 
closure after $m$ revolutions in $\theta_1$ and $l$ revolutions in 
$\theta_2$.
It is therefore appropriate to state that the phase space of a 2 
DOF system is structured by its periodic orbits.
In a 3 DOF system, however, motion on resonant tori is not 
generally closed, since the single condition ${\bf m}\cdot\omega = 0$ 
does not guarantee that any two of the three frequencies are commensurate.
The resonance condition guarantees only that the trajectory inhabits 
a submanifold of dimensionality two on the 3-torus; closure requires
the existence of an additional, independent resonance relation.
While resonant tori continue to lend phase space its structure, one 
can not assign every regular orbit in a 3 DOF system 
to a family associated with a particular periodic orbit.

\subsection{Numerical Recovery of the Fundamental Frequencies}

Recovering the fundamental frequencies of a regular orbit 
is an aspect of ``torus construction,'' the derivation of the map that 
relates Cartesian and action-angle variables.
Two general approaches to torus construction have been worked out 
in recent years.
Iterative techniques (\cite{cgm76}; \cite{rcs84}; \cite{mcb90}; 
\cite{war91}; \cite{bik93}; \cite{kab94}) begin by specifying the 
frequencies or actions of an orbit and then attempt to construct the 
relations ${\bf x}(\theta)$ that must be satisfied if the $\theta_i$ 
are to increase linearly with time.
These methods are often elegant but suffer from a lack of robustness: 
the equations to be solved are nonlinear and convergence 
can depend sensitively on the accuracy of the initial guess.
In a complex dynamical system containing many families of orbits, 
like the triaxial potentials considered here, 
it is generally necessary to ``fine-tune'' the algorithms in 
order to achieve convergence in each distinct phase-space region 
(e.g. \cite{kaa95}).

Trajectory-following algorithms (\cite{boo82}; \cite{kuo83})
avoid these problems by making explicit use of the quasi-periodic
nature of regular orbits.
Quasi-periodicity guarantees that the motion in any canonical
coordinate $x$ can be expressed as
\beq 
x(t) = \sum_{k=1}^{\infty}a_k e^{i\omega_kt} 
\label{quasip}
\eeq
where the $\omega_k$ are linear combinations of the fundamental 
frequencies, $\omega_k=l\omega_1+m\omega_2+n\omega_3$, and the $a_k$ are 
complex amplitudes.
By Fourier transforming the motion, the discrete peaks in the 
spectra can be identified as well as the corresponding 
amplitudes allowing a direct determination of $x(\theta)$.
The actions then follow from Percival's (1974) formula, 
completing the specification of the action-angle variables.
Binney \& Spergel (1982, 1984) pioneered this approach in the 
context of stellar dynamics, using a least squares technique to 
fit a model of the frequency spectra to $x(t)$.

Laskar's (1990) trajectory-following approach, adopted here, 
achieves much greater accuracy.
Laskar begins by replacing $x(t$) in equation (\ref{quasip}) 
by $f(t)$, some complex function of the orbit, e. g.
$f(t) = x(t) + iv_x(t)$, which is similar to the function 
$\zeta(t)$ that appears in equation (\ref{zeta}).
A close approximation to $f(t)$,
\beq 
f^\prime(t) = \sum_{k=1}^{N} a_k^\prime e^{i\omega_k^\prime
t},
\eeq
is then obtained in the following way.
First, the time series $f(t)$ is translated to an interval $[-T/2,T/2]$
symmetric about the time origin. 
Next, a discrete Fourier transform is taken of $f$ and the 
locations of the peaks $\omega_k'$ are identified.
The position of any peak will be defined to an accuracy of $\sim 
1/MT$ where $M$ is the number of equally-spaced intervals at 
which $f$ is recorded.
The estimate of the location strongest peak is then refined by 
computing the maximum of the function
\beq 
\phi(\omega) = \langle f(t), e^{i\omega t}\rangle = 
{1\over{T}}\int_{-T/2}^{T/2}{f(t)e^{-i\omega t}\chi(t)dt} 
\eeq 
where $\chi(t) = 1 + \cos({{2\pi t}/{T}})$ is the Hanning window 
function.  
The integral is approximated by interpolating the 
discretely-sampled $f$.
The Hanning filter broadens the peak but greatly reduces the 
sidelobes, allowing a very precise determination of $\omega_1'$.
Once the first frequency component has been found, its complex amplitude 
is obtained by projecting 
$e^{i\omega_1^\prime t}$ onto the original function $f(t)$, $a_1^\prime
= \langle f(t),e^{i\omega_1^\prime t}\rangle$.  
The first term in the
series is then subtracted, $f_1(t) = f(t) - a_1^\prime
e^{i\omega_1^\prime t}$ and the process repeated on $f_1(t)$ to find
the second frequency $\omega_2^{\prime}$. 
In general, the functions ${\bf e_1} = e^{i\omega_1^\prime t}$
and $e^{i\omega_2^\prime t}$ will not be orthonormal basis 
functions, but they can be made so by applying
a Gram-Schmidt orthogonalization process to obtain ${\bf e_2}$ 
of the form ${\bf e_2} = b_1 e^{i\omega_2^\prime t} - b_2
{\bf e_1}$, with $b_1$ and $b_2$ constants. 
${\bf e_2}$ is then projected onto $f_1(t)$ to obtain the corresponding 
amplitude $a_2'$. 
This process is repeated until the residual function does 
not signficantly decrease following subtraction of another term.
Typically, an accurate estimate of $f(t)$ is obtained with 
about 30 terms. 
We have found that extracting a larger number of terms from the
function, while possible, does not always improve the accuracy with
which the function $f(t)$ is estimated. 

The frequency analysis of $f(t)$ yields a set of 
frequencies $\omega_k$ which are linear combinations 
of the fundamental frequencies $(\omega_1, \omega_2, \omega_3)$.
Inferring the fundamental frequencies from the $\omega_k$ is an 
ill-defined problem, however, since any linearly-independent combination 
of the fundamental frequencies can equally well be interpreted as 
``fundamental.''
Suppose that a particular spike in the frequency spectrum has
$\omega_k = l\omega_1 + m\omega_2 + n\omega_3$. 
One can define new ``fundamental frequencies'' $\omega_i'$ as
\begin{eqnarray}
\omega_1 &=& n_{11}\omega_1' + n_{12}\omega_2' + n_{13}\omega_3',
\nonumber \\
\omega_2 &=& n_{21}\omega_1' + n_{22}\omega_2' + n_{23}\omega_3',
\nonumber \\
\omega_3 &=& n_{31}\omega_1' + n_{32}\omega_2' + n_{33}\omega_3'.
\nonumber
\end{eqnarray}
In terms of these new frequencies, $\omega_k$ becomes
\begin{eqnarray}
\omega_k&=&(ln_{11} + mn_{21} + nn_{31})\omega_1' + \nonumber \\
        & & (ln_{12} + mn_{22} + nn_{32})\omega_2' + \nonumber \\
        & & (ln_{13} + mn_{23} + nn_{33})\omega_3' \nonumber \\
        &=&l'\omega_1' + m'\omega_2' + n'\omega_3'. \nonumber
\end{eqnarray}
In practice, one expects that the largest amplitude term in each
frequency spectrum will correspond to a fundamental frequency.
However, this is only guaranteed to be true if the coordinate in 
question is close to an angle variable.

This ambiguity turns out not to be serious in the case of boxlike orbits, 
which can be thought of as perturbed rectilinear orbits.  
The fundamental
frequencies of a box orbit correspond approximately to motion in the
$x$, $y$ and $z$ directions, and so a frequency analysis of the motion
expressed in Cartesian variables yields highest-amplitude terms that
can almost always be identified with fundamental frequencies.  
The same is not true for the tube orbits.  
For instance, the
highest-amplitude term in both the $x$- and $y$-spectra of a short
($z$-) axis tube orbit is typically associated with the frequency of
rotation around the $z$ axis, $\omega_{\phi}$.  The term of next
highest amplitude typically has a frequency of
$\omega_R-\omega_{\phi}$, with $\omega_R$ the fundamental frequency
associated with motion in the radial direction.  

Despite these complications, we have successfully used an 
automated, integer-programming ``branch and bound'' algorithm from the NAG 
library (H02BBF) to extract the fundamental frequencies from 
the frequency spectra.
The entire set of frequencies $\omega_k$ is first sorted in
descending order of their real amplitudes. 
The frequency component of highest amplitude is defined to be the 
first fundamental frequency $\omega_1$.  
The entire series is then searched in descending order of
amplitude to find the second fundamental frequency $\omega_2$,
defined as the first term in the series which is not a simple integer
multiple of $\omega_1$.  
The integer programming routine then defines as $\omega_3$ the
next frequency component that is not a linear combination of
$\omega_1$ and $\omega_2$. 
In the case of box orbits and most tube orbits, the error in 
identifying the fundamental frequencies, ${\rm err} (\omega_k) =
|\omega_k-(l_k\omega_1+n_k\omega_2+m_k\omega_3)|$, is less than 
about $10^{-5}$.
Papaphilippou \& Laskar (1996) showed that polar coordinates were 
better suited than Carestian coordinates to obtaining the fundamental 
frequencies of tube orbits in an unambiguous way.
However, we found that an analysis of the tube orbits in terms of Cartesian
coordinates was satisfactory.  Two of the ``fundamental frequencies''
of a tube orbit returned by the integer programming routine could
almost always be interpreted as $\omega_{\phi}$ and
$\omega_{R}-\omega_{\phi}$, as discussed above; the third frequency 
corresponded to motion in a direction parallel to the symmetry axis of the tube.

Once the fundamental frequencies are obtained, one can define the
frequency map, a map from initial condition space to the space of
fundamental frequencies.  Since all orbits from a given ensemble were
selected to have the same energy, the distribution of points in
frequency space defines a two-dimensional surface; we follow
the practice of Laskar and co-authors of plotting $\omega_1/\omega_3$
vs.  $\omega_2/\omega_3$.  As discussed above, regular regions in the
frequency map are often associated with resonances between the
fundamental frequencies, ${\bf m\cdot\omega} = 0$.  The
highest-amplitude terms in the frequency spectrum of an orbit near a
stable resonance typically reflect the resonance condition.  For
instance, near the $x-z$ ``banana'' orbit, which has
$\omega_x/\omega_z=1/2$, one typically finds $\omega_1/\omega_3=0.5$
with very high accuracy.  An additional frequency, associated with the
slow libration around the resonance, is also present in the spectrum
but is generally small ($<10^{-6}\times \omega_1$) and of such low
amplitude that it is not selected by the integer-programming routine.
We again follow the practice of Papaphilippou \& Laskar (1996, 1998) of
defining the fundamental frequencies in the vicinity of a resonance as
the frequencies corresponding to the highest-amplitude terms.  Thus,
at least two of the fundamental frequencies of orbits in an island
around a boxlet will remain in exact proportion as one moves across
the island; the frequency corresponding to slow libration is ignored.
This practice results in the set of orbits around a stable resonance
mapping into a line in the frequency map, a very convenient
representation.

Stochastic orbits are not quasi-periodic and hence their 
frequency spectra can not be interpreted in terms of just three 
fundamental frequencies.
However, if an orbit is only weakly stochastic, the integer 
programming routine will still yield a set of ``fundamental 
frequencies'' in terms of which the various peaks in the 
frequency spectrum can be expressed with more or less accuracy.
As the degree of stochasticity increases, these ``fundamental 
frequencies'' become simply the three 
$\omega_k$ corresponding to the three strongest peaks in the 
frequency spectrum.
So defined, the ``fundamental frequencies'' of a stochastic orbit will 
depend on the time interval over which the orbit was integrated.
By integrating a stochastic orbit over two successive
time intervals $[0,T], [T, 2T]$, one can compute the change in 
these frequencies, which is a measure of the rate of diffusion 
in frequency space (\cite{las93}). 

In the remainder of the paper we will refer to our algorithm 
for carrying out the frequency analysis and integer
programming as ``NAFF,'' even though it differs slightly from
Laskar's algorithm.

\subsection{Accuracy of the NAFF Algorithm}

In this section, we discuss some of the factors that affect the accuracy
with which the fundamental frequencies of an orbit can be recovered.

Orbits were integrated using an explicit Runge-Kutta scheme of order 8
based on the method of Prince \& Dormand (1981).  
The program DOP853 from Hairer, Norsett \& Wanner (1993) uses local 
error estimation and adaptive step size control based on embedded 
formulas of orders 5 and 3 respectively.  
An important feature of the code is its ability to
produce ``dense output,'' accurate estimates of the dependent
variables at time intervals much shorter than the
integration time step.  The program DOP853 produces dense output via
7th-order interpolation between the output points.  The dense output
option greatly speeded up the construction of the time series needed
for the frequency analysis routine.

In order to estimate the accuracy of the numerically-computed 
$\omega_i$, Laskar (1993) suggested the following bootstrap 
scheme.
Call $f^{\prime}(t)$ the numerical approximation of the original 
time series $f(t)$, as in equation (11).
Carry out a second frequency analysis of $f^{\prime}(t)$.
The result is
\beq
f^{\prime\prime}(t) = 
\sum_{k=1}^{N} a_k^{\prime\prime}e^{i\omega_k^{\prime\prime}t}.
\eeq
Then $\delta a_k = |a^{\prime\prime}_k-a^{\prime}_k|$ and
 $\delta \omega_k = |\omega^{\prime\prime}_k-\omega^{\prime}_k|$ 
are estimates of the precision with which $a_k^{\prime}$ and
$\omega_k^{\prime}$ have been determined.

\begin{figure}
\vspace{7.cm}
\includegraphics{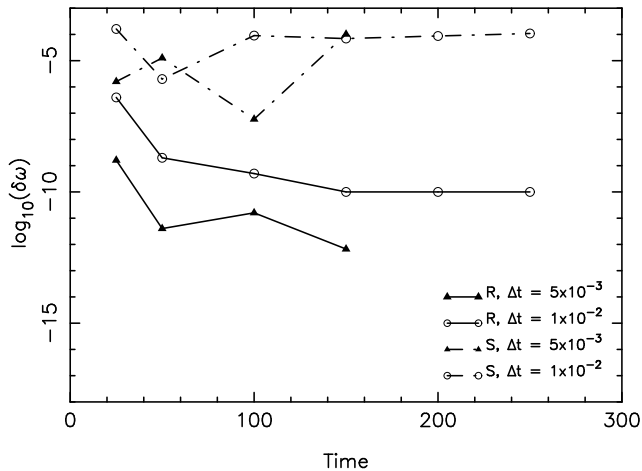}
\caption[]{
Dependence of Laskar's bootstrap accuracy estimate
$\delta\omega = |\omega^{\prime\prime}-\omega^{\prime}|$
on the orbital integration time.
The solid lines are for a regular orbit, with the time series
sampled at two different intervals $\Delta t$ as indicated.
The broken lines are for a stochastic orbit.
}  \label{fig1}
\end{figure}

Laskar (1996) has shown that frequency analysis with a Hanning filter
yields estimates of the fundamental frequencies whose accuracies scale
asymptotically as $1/T^4$, with $T$ the integration interval.  For an
ordinary FFT the dependence is only $1/T$.  Two additional factors
affecting the accuracy of the recovered frequencies are the sampling
interval $\Delta t$ and the accuracy of the numerical
integrator. 
We estimated the accuracy of the NAFF routine by applying
it to two orbits in a $\gamma$-model with $c/a = 0.5$ and $\gamma
= 0.5$.  Figure \ref{fig1} shows how the error in the fundamental
frequencies of two different orbits varies with $T$ and $\Delta t$;
the time unit is the period of the long-axis orbit of the same energy.
$\delta\omega$ was defined as $|\omega^{\prime\prime} -
\omega^{\prime}|$ for the frequency component of the largest 
amplitude.  The solid lines were obtained for a regular orbit, and the
broken lines were obtained for a stochastic orbit.  The open circles
correspond to a larger sampling interval of $\Delta t \sim 10^{-2}$
($\sim 100$ steps per orbital period) and the filled circles
correspond to a smaller sampling interval of $\Delta t
\sim 5\times 10^{-3}$ ($\sim 200$ steps per orbital period). 
This figure illustrates the reduction in $\delta\omega$ for a regular
orbit as the integration interval is increased.  The smaller sampling
interval also improves the bootstrap estimate of the accuracy $\delta
\omega$ by nearly two orders of magnitude.  In the rest of the paper
we adopt an integration interval of 50 orbital periods (defined as the
period of the long-axis orbit of the same energy) and a sampling
interval $\Delta t = 5 \times 10^{-3}$.  Figure \ref{fig1} also shows
that the ``frequencies'' of a stochastic orbit are obtained with a much
lower accuracy ($10^{-4}-10^{-6}$) than for a regular orbit.  Clearly,
the function $f^{\prime}(t)$ obtained for a stochastic orbit is not a
good approximation to the original $f(t)$.

\begin{figure}
\vspace{7.cm}
\includegraphics{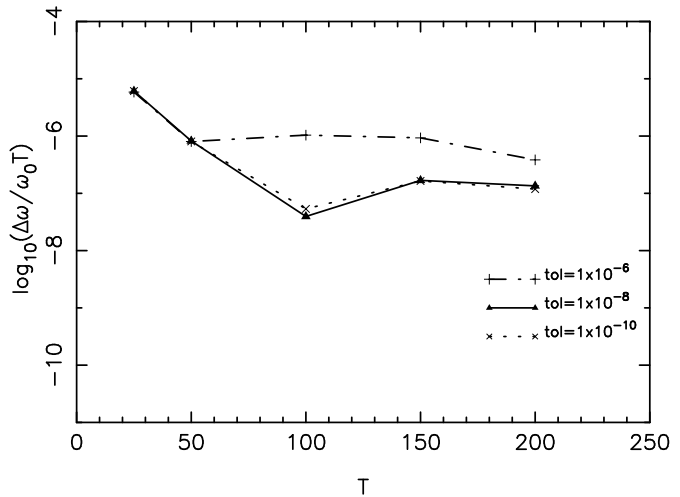}
\caption[]{
Diffusion rate $\Delta \omega$  obtained by measuring the
frequencies of a regular orbit over two successive time intervals
of length $T$, plotted as a function of $T$, for three different values
of the tolerance parameter of the orbit integrator.
}
\label{fig2}
\end{figure}

Another way to estimate the accuracy of the NAFF routine is to 
integrate a single, regular orbit over two successive time intervals and 
compare the estimates of the fundamental frequencies.
We define $\Delta \omega = |\omega(T_1)-\omega(T_2)|/\omega_0T$, 
where $T_1$ and $T_2$ are the midpoints of the two intervals, 
$T=T_2-T_1$, and $\omega_0=2\pi/T_X$, the frequency of the long-axis
orbit of the same energy.
Figure \ref{fig2} shows $\Delta\omega$ as a function of $T$ and 
of the tolerance parameter of the numerical integrator for the 
regular orbit of Figure \ref{fig1}.
Figure \ref{fig2} indicates that
decreasing the tolerance parameter below $1\times 10^{-8}$
does not significantly affect the value of $\Delta \omega$ when
the sampling interval is fixed to this value.
Unless otherwise indicated, results in this paper were
obtained with a tolerance parameter of $10^{-8}$.

\section{The Structure of Phase Space}

\begin{figure*}
\vspace{9.5cm}
\includegraphics{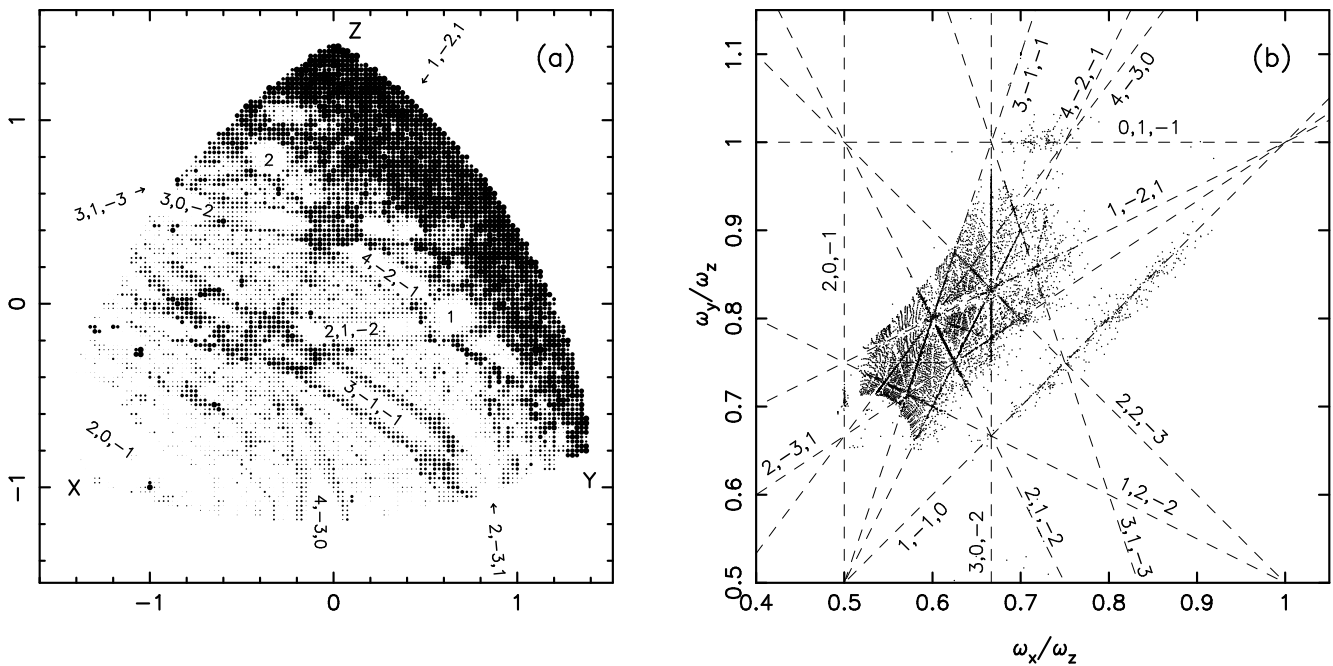}
\caption[]{
Box-orbit phase space at shell 8 in the model with $\gamma=0.5$.
(a) Initial condition space: one octant of the equipotential surface 
has been projected onto a plane.
Each orbit begins on this surface with zero velocity.
The top, left and right corners correspond to the $z$ (short), 
$x$ (long) and $y$ (intermediate) axes.
The grey scale is proportional to the logarithm of the 
diffusion rate of orbits in frequency space; initial conditions 
corresponding to regular orbits are white.
The regions labelled ``1'' and ``2'' contain orbits associated 
with the $5:6:8$ and $7:9:10$ periodic orbits, respectively.
Other regions are labelled with the integers $(l,m,n)$ that 
define resonance zones.
(b) Frequency map: the fundamental frequencies returned by the 
NAFF algorithm are plotted as rotation numbers
$\omega_x/\omega_z$ and $\omega_y/\omega_z$ for each of the 
orbits in (a).
The most important resonances are labelled.
Stable resonances produce solid lines; gaps correspond 
to unstable resonances.
}  \label{fig_map1}
\end{figure*}

We begin by looking in detail at a triaxial $\gamma$-model 
with $c/a=0.5$ and $\gamma=0.5$, a ``weak cusp.''
Two ensembles of $\sim 10^4$ orbits were integrated in this model 
at the fixed energy corresponding to shell 8, which lies just inside the 
half-mass radius.
Initial conditions for the two ensembles were chosen so as to 
generate either boxlike orbits or tube orbits; i.e. orbits with 
stationary points, or orbits that circulate about one of the 
principal axes of the model.
(A more complete discussion of the two initial-condition spaces may
be found in Merritt \& Fridman 1996).
Orbits were integrated for a total of 100 periods of the $x$-axis 
orbit.
Fundamental frequencies were computed independently for 
the first and second halves of this interval so that diffusion 
rates could be calculated, as described above.

\subsection{Box-Orbit Initial Conditions}

A large part of the phase space of a triaxial system is associated
with orbits that have a stationary point, i.e. that touch the
equipotential surface with zero velocity.  All of the box orbits in an
integrable (St\"ackel) potential fall into this category, as do the
``centrophobic'' boxlets, like the $x-z$ banana, in non-integrable
potentials.  
(``Centrophilic'' boxlets, like the anti-banana, typically do not 
have a stationary point.
Such orbits are generally unstable (\cite{mis89})).
In Figure \ref{fig_map1}, the
NAFF algorithm has been applied to a set of 9408 orbits whose initial
conditions lay in a regular grid on one octant of the equipotential
surface corresponding to shell 8.  Figure \ref{fig_map1}a presents a
planar projection of the initial condition space; the gray scale has
been adjusted in proportion to the logarithm of the diffusion rate,
defined as the change in the largest-amplitude fundamental frequency
as evaluated over the initial and final intervals of 50 periods each.
Figure \ref{fig_map1}b is the frequency map corresponding to the first
integration interval.  
In both figures, resonances defined by 
$l\omega_x+m\omega_y+n\omega_z=0$ 
have been identified for certain integer values of $(l,m,n)$.

The frequency map exhibits some regions in which the points are
arranged in a more-or-less orderly fashion.  
These regions may be
identified with parts of phase space that have retained their regular
character in spite of the perturbation of the potential away from 
integrable, or St\"ackel, form.  Separating these regular regions are
the resonance lines: either solid lines, corresponding to stable
resonant tori; or empty gaps, corresponding to stochastic layers.  
In addition, a number of points are scattered in a
more irregular manner around the figure.  These are the stochastic
orbits.  However, even the stochastic orbits are mostly clustered
around resonance lines, indicating that the stochastic motion is often
strongly influenced by structures in phase space that are similar to
invariant tori.

The short ($z-$), intermediate ($y-$) and long ($x-$) axis orbits are
all unstable to lateral perturbations at this energy in this potential
(\cite{frm97}).  Instability of the $z-$ axis orbit first occurs at
low energy through bifurcation of the $1:1$ $y-z$ loop orbit
(\cite{gos81}).  The influence of this resonance is visible at the top
of Figure \ref{fig_map1}b, where many of the stochastic orbits started
near the $z$ axis are clustered around the $(0,1,-1)$ resonance line.
(The stable branch of the bifurcation, the $y-z$ loop orbits, do not
appear in this plot since they have no stationary points.)
Instability of the $y-$ axis orbit first occurs
through the $1:1$ $x-y$ loop bifurcation; the region around the
$(1,-1,0)$ resonance is also well populated in the frequency map, by
stochastic orbits that start near the $y-$ axis.  Finally, the $x-$
axis orbit becomes unstable at the bifurcation of the $1:2$ $x-z$
banana orbit, which occurs at about shell 2 in this model
(\cite{frm97}).  Unlike the $1:1$ loop orbits, however, the stable
banana orbit does lie in stationary initial condition space, and hence
the family of stable orbits generated by the bifurcation appear in
the frequency map along the $(2,0,-1)$ resonance line.  A smaller
number of stochastic orbits lie near to, but offset from, the
$(2,0,-1)$ line.

As emphasized by Papaphilippou \& Laskar (1998), the motion in 
boxlike phase space is strongly influenced by resonances 
between the three degrees of freedom.
Such resonances are dense in the phase space of an unperturbed, i.e. 
integrable, Hamiltonian, in the sense that every torus lies near to a torus 
satisfying ${\bf m\cdot\omega}=0$ for some (perhaps very large) 
integer vector $\bf m$.
However, most tori are very non-resonant in the sense that 
${\bf m}\cdot\omega$ is large compared with $|{\bf m}|^{-(N+1)}$, 
with $N$ the number of degrees of freedom.
These very non-resonant tori persist under finite perturbations of 
the Hamiltonian and may be identified with the orderly regions 
between the resonance lines in the frequency map of Figure \ref{fig_map1}b.
Resonant tori, particularly those with small $\bf m$, are either 
destroyed by a finite perturbation, generating stochastic motion, 
or become associated with a regular region surrounding the 
resonance, or (typically) both.
Many -- perhaps most -- of the regular orbits in Figure \ref{fig_map1}a
can be identified with a stable resonance; among the most 
important are the $(2,1,-2)$, $(3,-1,-1)$, and $(4,-2,-1)$ resonances.
The intersection of any two resonance lines defines a periodic 
orbit, i.e. an orbit for which $\omega_x/\omega_z=l/n$ and 
$\omega_y/\omega_z=m/n$.
When stable, a periodic orbit can have its own associated 
regular region, distinct from the regular regions associated with 
the 3D resonances.
At least two such regions are apparent in Figure \ref{fig_map1}a, 
associated with the $5:6:8$ and $7:9:10$ periodic orbits (marked
1 and 2).
In addition, the $x-z$ fish ($2:3$) and $x-y$ pretzels ($3:4$) planar
boxlets have associated regular regions.

\begin{figure}
\vspace{10.5cm}
\includegraphics{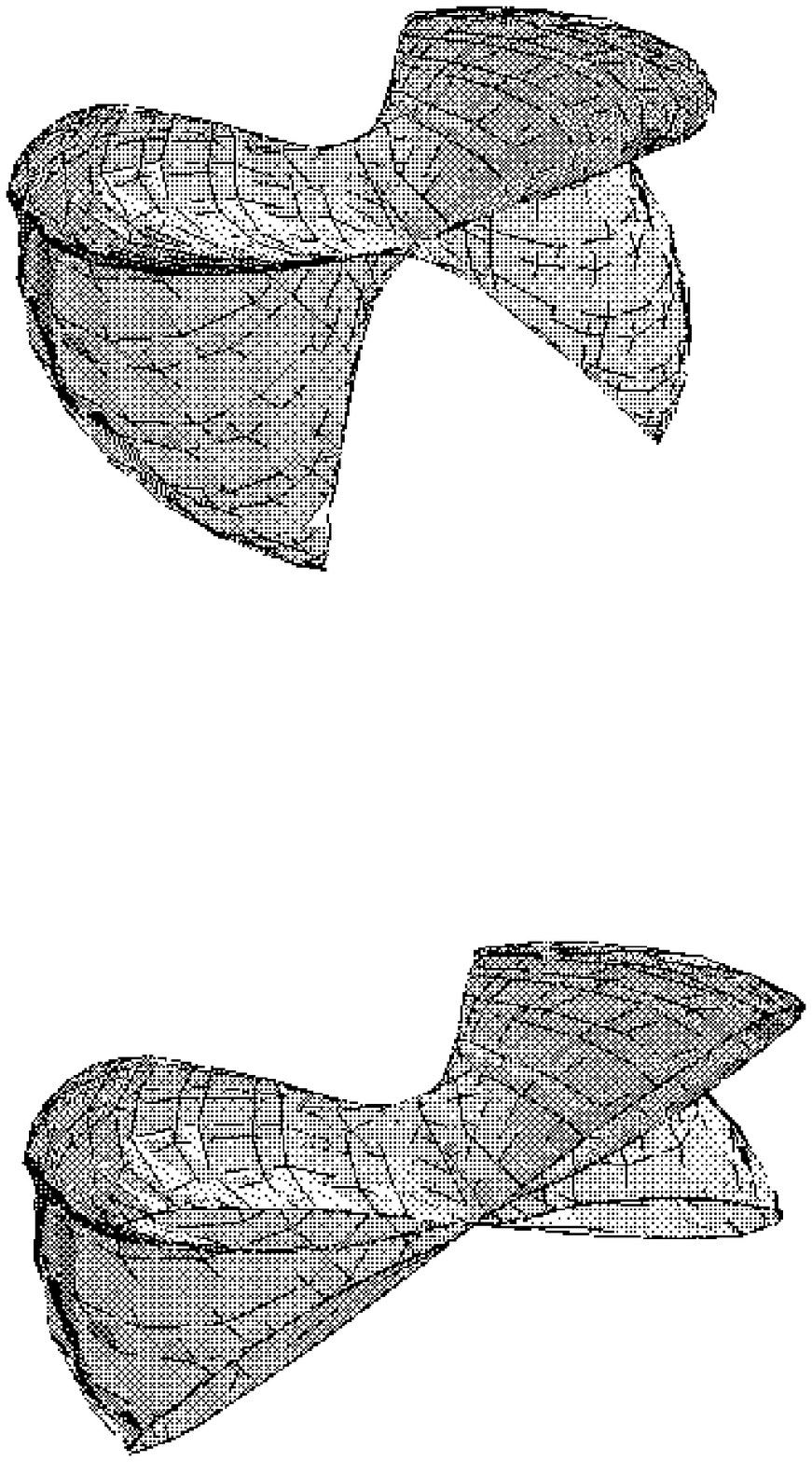}
\caption[]{
A thin box orbit associated with the $(2,1,-2)$ resonance.
The lower view is a cutaway showing that the orbit is confined
to a membrane.
}  \label{fig_thinbox}
\end{figure}

It is sometimes stated that all of the regular orbits in a galactic 
potential belong to families that can be associated with stable 
periodic orbits.
For instance, the box orbits in a St\"ackel potential are commonly 
associated with the long-axis orbit, while the tube orbits 
are said to belong to families generated by the $x-y$ or $y-z$ $1:1$ 
planar loops.
While the identification of regular orbits with stable periodic 
orbits is sometimes justified, only a small fraction of the regular 
orbits in Figure \ref{fig_map1}a can be usefully assigned to families 
associated with periodic orbits.
A larger fraction of the regular orbits are located in resonance 
{\it zones}, regions associated with tori that satisfy 
a single resonance condition $l\omega_x+m\omega_y+n\omega_z=0$.
We expect orbits in such regions to be approximately confined to 
two-dimensional sub-manifolds; in configuration space, they should 
take the form of thin, curved sheets.
An example of such an orbit, which we call a ``thin box,'' is shown
in Figure \ref{fig_thinbox}.
Closure requires an additional resonance condition
${\bf m\cdot\omega}=0$ to be satisfied so that the frequency vector
can be written as ${\bf\omega}={\bf n}\omega_0$ with ${\bf n}$ an
integer vector.

In initial condition space (Figure \ref{fig_map1}a), the regions associated 
with stable resonance zones appear as bands of white, flanked 
on both sides by narrower stochastic strips.
Periodic orbits lie at the intersection of two or more of these 
bands, and some of these periodic orbits -- e.g. the $5:6:8$ and 
$7:9:10$ orbits in Figure \ref{fig_map1}a -- are surrounded by regions in 
which the motion is regular.
Orbits in these restricted regions may be usefully associated with the 
periodic orbit at their center.
But Figure \ref{fig_map1}a suggests that the regular regions associated 
with stable periodic orbits are relatively small, and they are often
separated from the larger regular regions by strips where the motion is 
stochastic.
There would seem to be no reason to associate the regular motion in the 
largest part of any resonance zone with any particular periodic 
orbit.

In a 2 DOF system, the condition $l\omega_1+m\omega_2=0$ that 
defines a resonant torus also defines a periodic orbit, since
$\omega_1/\omega_2=-m/l$ implies commensurability of the motion in 
the two directions around the torus.
The distinction between periodic orbits and resonant tori 
therefore does not appear in systems with fewer than three degrees of 
freedom.
In the same way, it is clear from Figure \ref{fig_map1}a that orbits 
restricted to, say, the $x-y$ plane are affected by the 
$(4,-3,0)$ resonance zone only where that zone intersects the 
$x-y$ plane, at the $x-y$ pretzel orbit.
In the case of the majority of non-planar regular orbits, 
however, no unique identification with a periodic orbit is possible.

In addition to the regular orbits associated with resonance zones 
and with periodic orbits, a third class of regular orbit is
apparent in Figure \ref{fig_map1}.
These are orbits that lie in the (essentially) completely regular 
regions that lie between the resonance zones.
Orbits in these regions may be identified with very 
non-resonant tori, i.e. tori that have maintained their integrity
in spite the perturbation of the Hamiltonian away from integrable form.
In an integrable potential, all of the regular orbits would 
belong to such regions; in a non-integrable potential, only tori 
that are sufficiently non-resonant are expected to avoid being 
strongly deformed by the perturbation.
Strictly speaking, we expect to find some high-order 
resonance zones and associated stochastic regions imbedded within 
even these ``regular'' regions, and in fact a close inspection of 
Figure \ref{fig_map1}a reveals just such structure. 
However, the chaotic diffusion rates in these regions are 
generally small and for practical purposes the motion in them is 
effectively regular throughout.
Again, we note that the motion in one of these regular regions can not 
be properly identified with any single periodic orbit; instead, 
the regularity of the motion would seem to be properly identified with 
the persistence of integrability in extended, non-resonant parts of phase 
space in spite of the perturbation of the Hamiltonian away from fully 
integrable form.

\begin{figure*}
\vspace{9.75cm}
\includegraphics{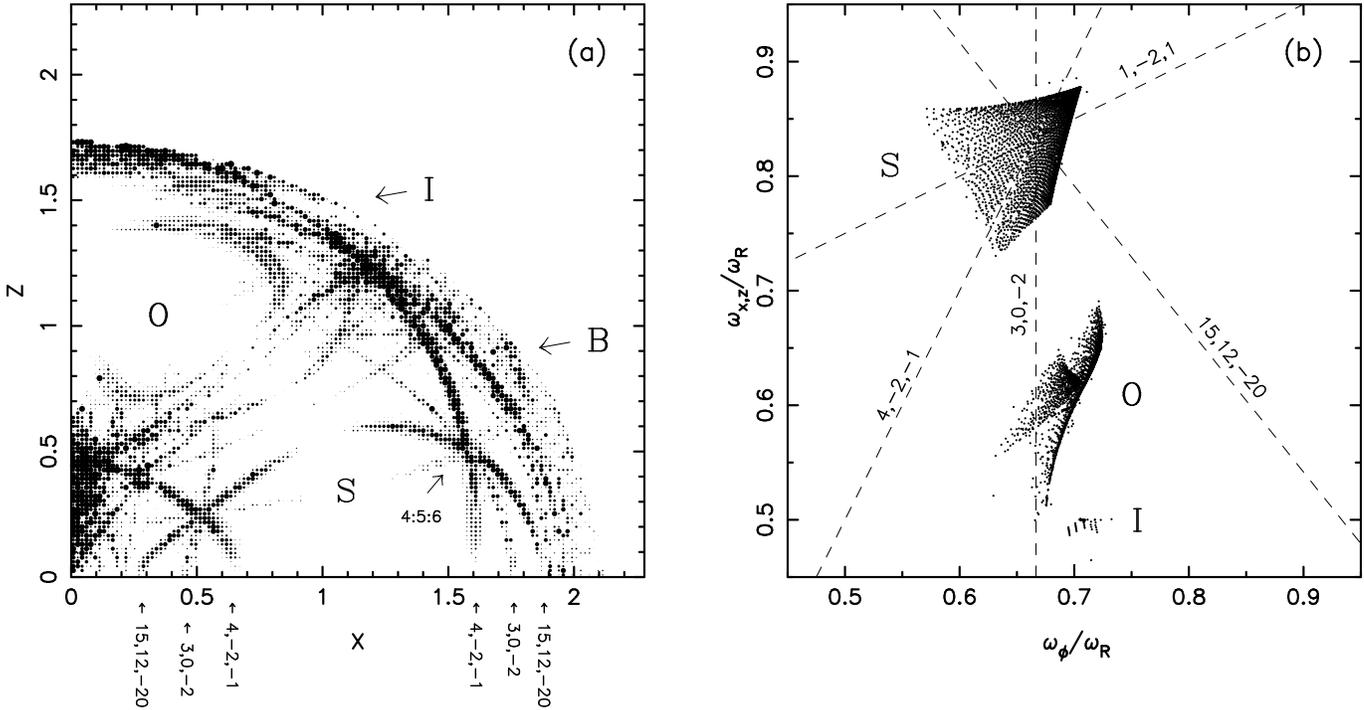}
\caption[]{
Tube-orbit phase space at shell 8 in the model with $\gamma=0.5$.
(a) Initial condition space: orbits begin on the $x-z$ plane with 
$v_x=v_z=0$.
The grey scale is proportional to the logarithm of the 
diffusion rate of orbits in frequency space; initial conditions 
corresponding to regular orbits are white.
The regions labelled ``S,'' ``O,'' ``I'' and ``B'' contain 
short-axis tubes, outer long-axis tubes, inner long-axis tubes 
and boxes, respectively.
The three most important resonance zones are labelled; they 
intersect at the (unstable) $4:5:6$ orbit (Figure \ref{fig_456}).
(b) Frequency map, with the important resonances labelled.
Only the tube orbits are included.
$\omega_{\phi}$ is the fundamental frequency associated with 
motion about the short axis (short-axis tubes), or long axis 
(long-axis tubes).
}
\label{fig_map2}
\end{figure*}

An equivalent way to define the three ``types'' of regular
orbit described above is in terms of the number of independent 
relations ${\bf m}\cdot\omega=0$ that characterize their associated 
phase space regions, i.e. in terms of the degeneracy of the resonance.
A periodic orbit is defined by two such relations; a resonance 
zone by one; and the regular regions that persist in spite of the
perturbation of the potential away from integrability are 
defined by none.
It is clear from this definition that the number of distinct ``types'' 
of regular orbit is equal to the number of degrees of freedom.
While regular orbits of all three types occupy three-dimensional 
regions in configuration space, orbits associated with a single 
resonance condition are approximately confined to sheets, while 
orbits associated with two resonance conditions are approximately 
confined to closed curves.

Given that essentially none of the boxlike orbits in Figure \ref{fig_map1}a 
can be properly identified with the long-axis orbit, it is reasonable to 
ask in what sense the box orbits in {\it any} triaxial potential -- even 
that of the  Perfect Ellipsoid -- are correctly said to belong to 
the ``family of orbits'' associated with the long-axis orbit.
We would argue instead that the regularity of the box orbits in 
a St\"ackel potential is properly ascribed to the global 
integrability of the potential, and not to the local stability of any 
single periodic orbit like the long-axis orbit.

The regions corresponding to stochastic motion in Figure \ref{fig_map1}a can 
likewise be separated into three fairly distinct types, with one important 
qualification.
Stochastic orbits live in a phase space of higher dimensionality than
regular orbits, and in fact we expect every stochastic orbit to 
eventually visit (via Arnold diffusion) every point of phase space, even 
points that are arbitrarily close to regular orbits.
However the timescale for Arnold diffusion is 
extremely long, and for practical purposes, stochastic orbits -- like 
regular orbits -- can often be associated with restricted parts 
of phase space.
Accordingly, in Figure \ref{fig_map1}a, we identify three ``types'' 
of stochastic orbit.
First are the stochastic orbits closely associated with an 
unstable periodic orbit.
An example is the region around the $4:5:6$ periodic orbit that 
lies at the intersection of the $(3,0,-2)$ and $(4,-2,-1)$ 
resonance zones.
Second, some stochastic orbits lie along the edges of the stable 
resonance zones, as mentioned above. 
Third, there is a large stochastic strip connecting the $y-$ and 
$z-$axis orbits.
Goodman \& Schwarzschild (1981) first described this stochastic 
strip in their study of stochastic motion in a triaxial model with a core.
The overlap of the chaotic zones associated with the $y$ and $z$-axis 
orbits appears to be responsible for most of this strip, although a 
number of other resonances are also important, including especially 
the $(1,-2,1)$ resonance.
Although the diffusion rate of orbits in the stochastic strip is 
high, we would not expect these orbits to wander ergodically over 
the entire energy surface due to the existence of substantial regular 
regions which would inhibit the diffusion.
However, the near constancy of the diffusion rate throughout the 
$y-z$ strip suggests that the motion is nearly ergodic over this 
restricted region, at least for elapsed times of $\sim 10^2$ 
oscillations or more.

\subsection{Tube-Orbit Initial Conditions}

The other major category of orbits in a triaxial potential, the tube 
orbits, are characterized by a finite, time-averaged angular momentum 
about the long or short axis.
In an integrable triaxial potential, the three families of tube 
orbits -- the short-axis tubes, the inner long-axis tubes, and 
the outer, long-axis tubes -- can all be recovered by taking initial 
conditions in the $x-z$ plane, with $v_y=0$ (\cite{sch93}).
This initial condition space will also include some box orbits.
Figure \ref{fig_map2}a presents the $x-z$ initial condition space for the same 
model of Figure \ref{fig_map1}; again, the grey scale is adjusted in 
proportion to the logarithm of the diffusion rate.
The corresponding frequency map is shown in Figure \ref{fig_map2}b.
Any non-periodic orbit intersects the $x-z$ plane in at least two points; 
in the case of tube orbits, the two points lie inside and outside 
the curve that defines the corresponding ``thin'' orbit family.

As discussed above, the highest-amplitude components in the $x-$ and $y-$ 
spectra of a short ($z-$) axis tube are typically both associated with the 
frequency of rotation about the $z$ axis, $\omega_{\phi}$; the 
frequency corresponding to radial oscillations, $\omega_R$, 
appears typically in terms of much lower amplitude.
For a few of the $\sim 10^4$ tube orbits integrated here, 
the automated routine for extracting fundamental frequencies 
appeared to fail, leaving some random gaps in the frequency map 
of Figure \ref{fig_map2}b.
(Papaphilippou \& Laskar (1996) reported similar difficulties in 
their analysis of tube orbits.)
However, the fundamental frequencies of the great majority of regular tube 
orbits were successfully recovered.
In Figure \ref{fig_map2}b, only orbits that clearly belonged to one of the 
three tube families are plotted in the frequency map; the boxlike 
orbits are omitted for clarity.
The fundamental frequency labeled $\omega_{\phi}$ is defined as 
the frequency associated with motion around the $z$ axis in the 
case of the short-axis tubes (labeled ``S'' in the figure), and 
as the frequency associated with motion around the $x$ axis in 
the case of the inner (I) and outer (O), long-axis tubes.

\begin{figure}
\vspace{7.5cm}
\includegraphics{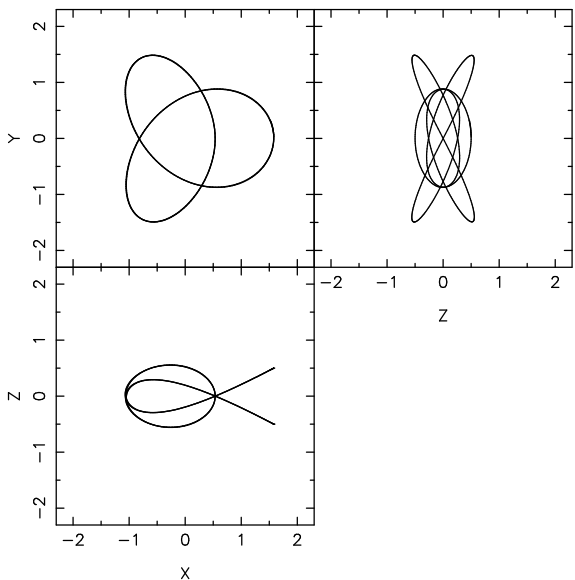}
\caption[]{
The unstable, $4:5:6$ resonant tube orbit that appears in Figure \ref{fig_map2}}.
\label{fig_456}
\end{figure}

This part of phase space is largely regular.
Aside from the box orbits, which are confined to regions near the 
zero-velocity curve, all stochastic tube orbits are associated with 
a handful of narrow resonance zones.
These zones exist at the transition points between the various 
families, as noted by Merritt \& Fridman (1996) and Papaphilippou 
\& Laskar (1998).
A few additional resonances are important, including 
the $(4,-2,-1)$, $(3,0,-2)$ and the surprisingly high-order 
$(15,12,-20)$ resonance.
These three resonances intersect at the $4:5:6$ 
periodic orbit, an unstable tubelet (Figure \ref{fig_456}).
The diffusion rates appear to drop to zero along the curve 
defining the thin tube orbits belonging to the $S$ and $O$ 
families, a reasonable result.

\begin{figure}
\vspace{7.5cm}
\includegraphics{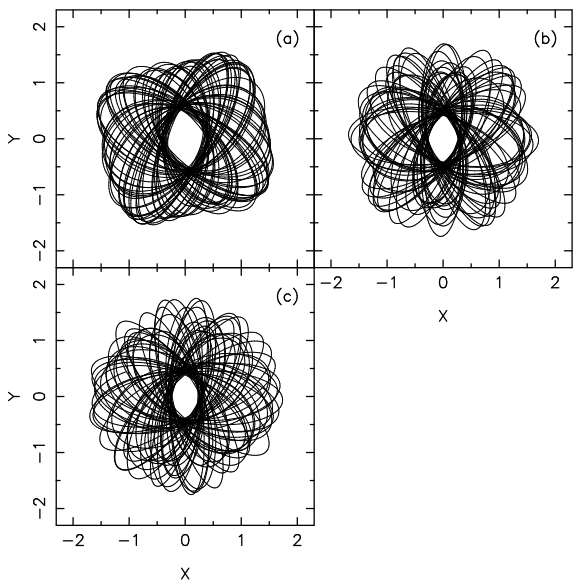}
\caption[]{
A stochastic, short-axis tube orbit.
The initial conditions lie in Figure \ref{fig_map2}a on the resonance zone 
labelled $(4,-2,1)$, slightly above the $4:5:6$ periodic orbit.
(a), (b) and (c) show three successive integrations of the orbit, 
each over a time span of 50 periods of the long-axis orbit.
}
\label{fig_ctube}
\end{figure}

We have encountered little discussion in the literature 
of stochastic tube orbits; the nearest examples that we know of
are orbits that Kandrup, Eckstein \& Bradley (1997) describe as appearing
``alternately boxy and loopy.''  
Accordingly, we illustrate in Figure \ref{fig_ctube} 
the evolution of a tube orbit 
associated with the unstable $(4,-2,-1)$ resonance zone in Figure 
\ref{fig_map2}a.
The orbit was integrated for three successive intervals equal to 
$50$ periods of the long-axis orbit.
The stochasticity is evident in the gradual change of the
orbit's shape.
Assuming that the entire stochastic phase space at this energy is
interconnected, we would expect this tube orbit to eventually 
find its way into boxlike phase space.
However the time required is likely to be extremely long, since 
the diffusion would have to occur along the $(4,-2,-1)$ resonance 
line, and diffusion along resonance layers -- i.e. Arnold diffusion -- is 
known to be extremely slow.
\section{The Transition to Global Stochasticity}

When an integrable potential is perturbed by increasingly large
amounts, the fraction of phase space associated with stochastic motion
and the rate of stochastic diffusion typically increase.
In triaxial $\gamma-$models, the parameter $\gamma$ that
defines the steepness of the central density cusp is expected to
play the role of a perturbation parameter: 
models with $\gamma$ close to zero should be
close to integrable (though never exactly so, since the $\gamma$-models 
do not reduce to integrable form even for $\gamma=0$), 
while values of $\gamma$ greater than about one are known to generate 
strongly stochastic motion (\cite{mef96}).  
The mass of a central singularity, and the energy, should also act as 
perturbation parameters; for instance, the long-axis orbit generally becomes
unstable at sufficiently high energies even in triaxial potentials
with cores.
In terms of any of these perturbation parameters,
boxlike phase space should be most strongly affected  
since tube orbits avoid the destabilizing center.

In the case of extreme perturbations, dynamical systems sometimes
exhibit a transition to global stochasticity, a regime in which the stochastic 
motion is interconnected over large portions of the phase space. 
(Strictly speaking, stochastic phase space in a 3 DOF system is always 
interconnected through the Arnold web, but Arnold diffusion is 
extremely slow unless the stochastic regions overlap.)
In the globally stochastic regime, there are few barriers to the 
motion and orbits can wander over the energy 
hypersurface in little more than an orbital period.
One standard definition of the transition to global stochasticity 
is based on the overlap of the stochastic layers 
associated with the primary resonances (\cite{chi60}).
Other definitions (\cite{lil92}) are based on the fraction 
of phase space that is stochastic, the degree to which 
the stochastic regions are interconnected, etc.

In triaxial stellar systems, a transition to global 
stochasticity implies the replacement of distinguishable box orbits by
orbits that move nearly ergodically over the energy surface and 
that densely fill the configuration space volume defined by an 
equipotential surface.
Since box orbits with various shapes are always found to be strongly 
populated in self-consistent triaxial models (\cite{sch79}; \cite{sta87}), 
significant triaxiality 
should be difficult or impossible to maintain in a system where the 
motion is globally stochastic.
This hypothesis has received some support from self-consistent 
equilibrium studies (\cite{sch93}; \cite{mer97}) and from 
$N$-body integrations (\cite{meq98}).

We investigated the approach to global stochasticity in the triaxial
$\gamma$-models by applying the NAFF algorithm to a set of models with 
different values of the central density slope $\gamma$ and
with central point masses $M_h$ of various sizes.
(Dependence of the degree of stochasticity on the energy is 
discussed in the following section.)
In each model, we integrated ensembles of $\sim 10^4$ orbits at 
shell 8 in boxlike initial condition space and computed their 
fundamental frequencies over two adjacent time intervals with the NAFF 
algorithm.
We carried out similar experiments in the initial condition 
space of tube orbits; however, as expected, the degree of stochasticity was 
not found to depend strongly on the central density structure and 
so those results will not be presented here.

\begin{figure*}
\vspace{21cm}
\includegraphics{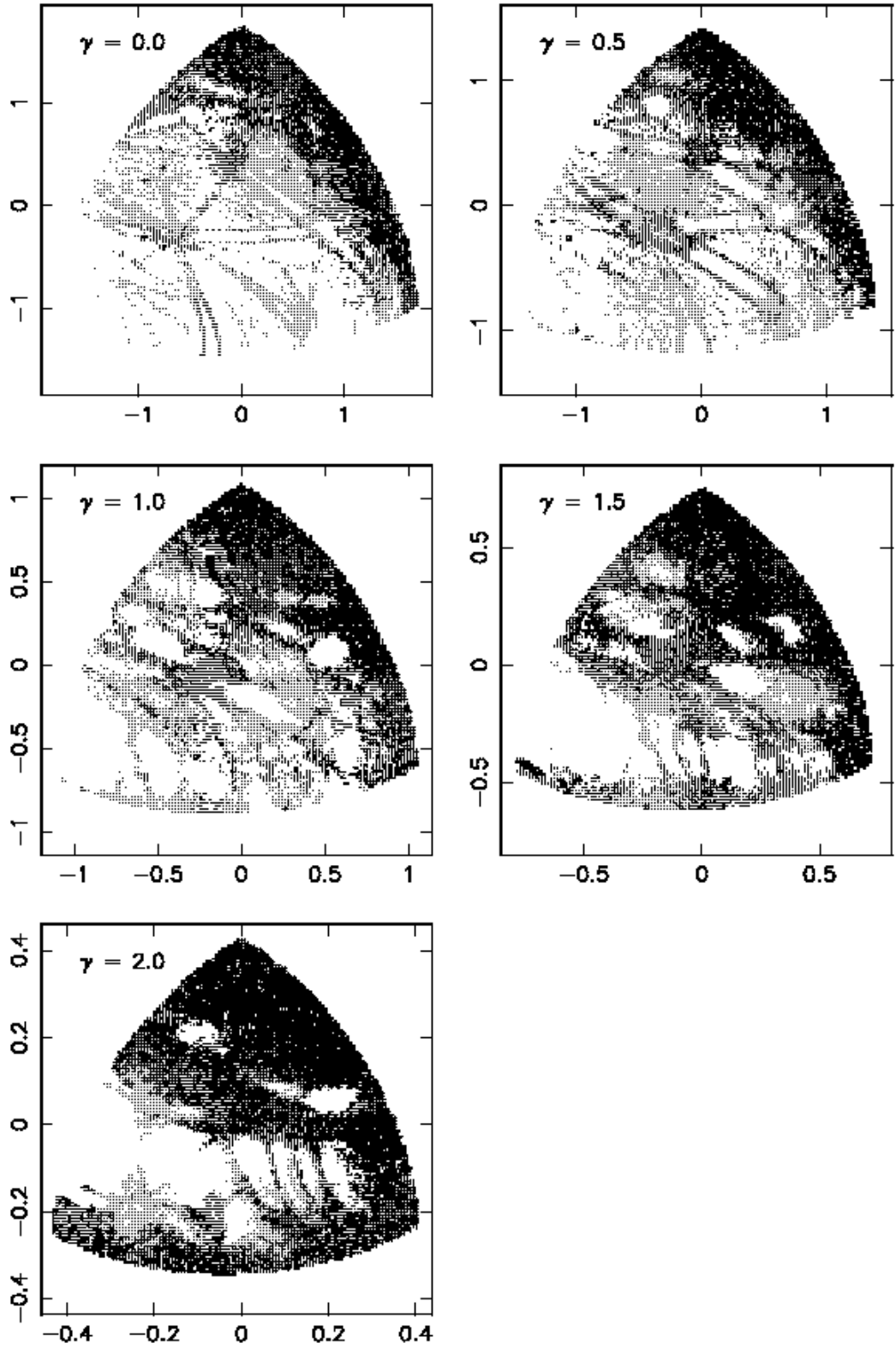}
\caption[]{
Initial condition space at shell 8 in a set of models with 
various values of the cusp slope $\gamma$.
The grey scale is proportional to the logarithm of the diffusion 
rate in frequency space, defined as the fractional change in the 
largest-amplitude fundamental frequency over 50 periods of the 
long-axis orbit.
White regions correspond to regular orbits.
}
\label{fig_diff1}
\end{figure*}

\begin{figure*}
\vspace{21cm}
\includegraphics{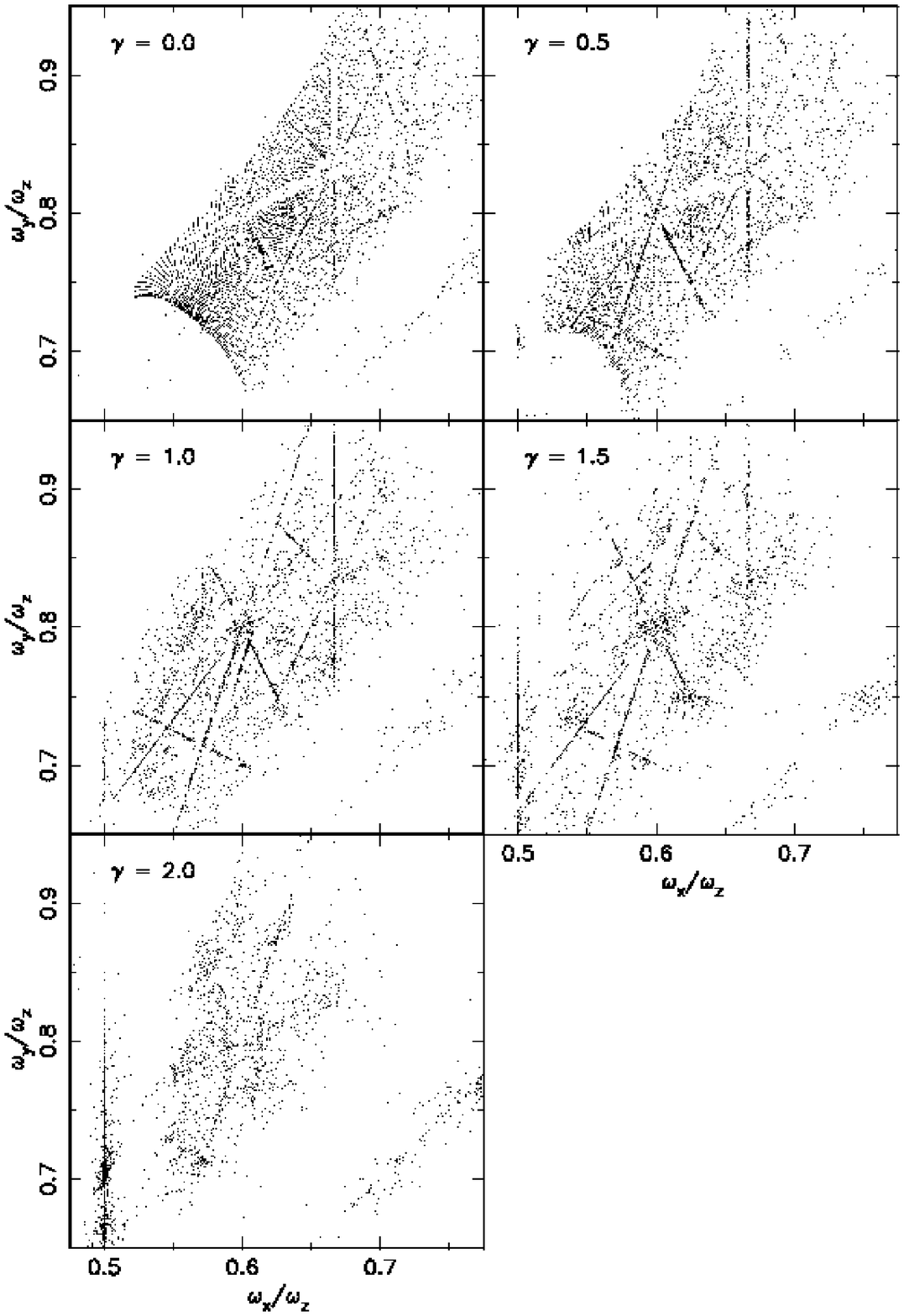}
\caption[]{
Frequency maps for the orbits of Figure \ref{fig_diff1}.
}
\label{fig_fm_1}
\end{figure*}

\begin{figure*}
\vspace{21cm}
\includegraphics{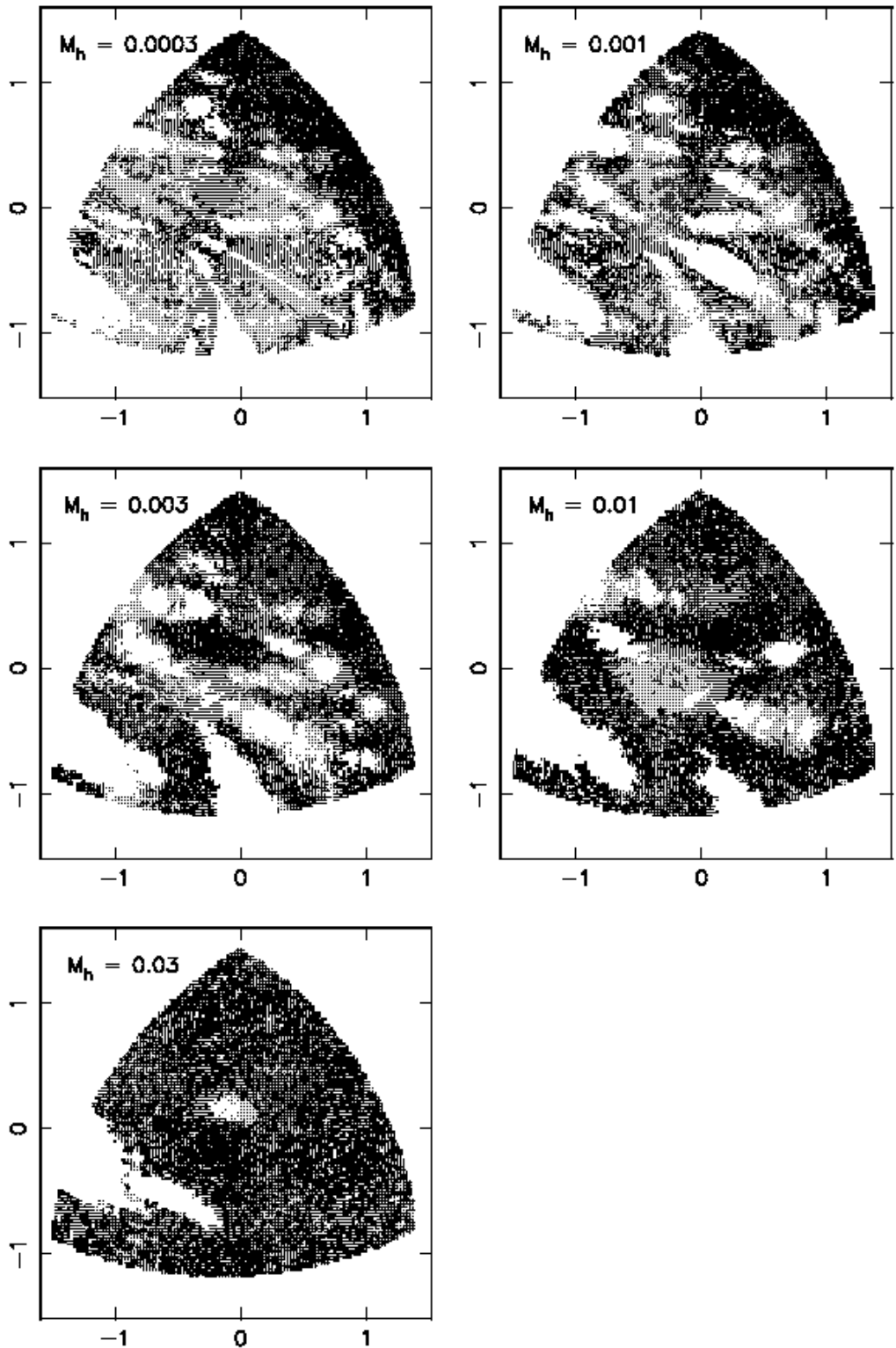}
\caption[]{
Initial condition space at shell 8 in a set of models with 
$c/a=T=\gamma=0.5$, and various values of $M_h$, the mass of a 
central singularity.
The grey scale is proportional to the logarithm of the diffusion 
rate in frequency space, defined as the fractional change in the 
largest-amplitude fundamental frequency over 50 periods of the 
long-axis orbit (in the model without a central point mass).
White regions correspond to regular orbits.
}
\label{fig_diff2}
\end{figure*}

\begin{figure*}
\vspace{21cm}
\includegraphics{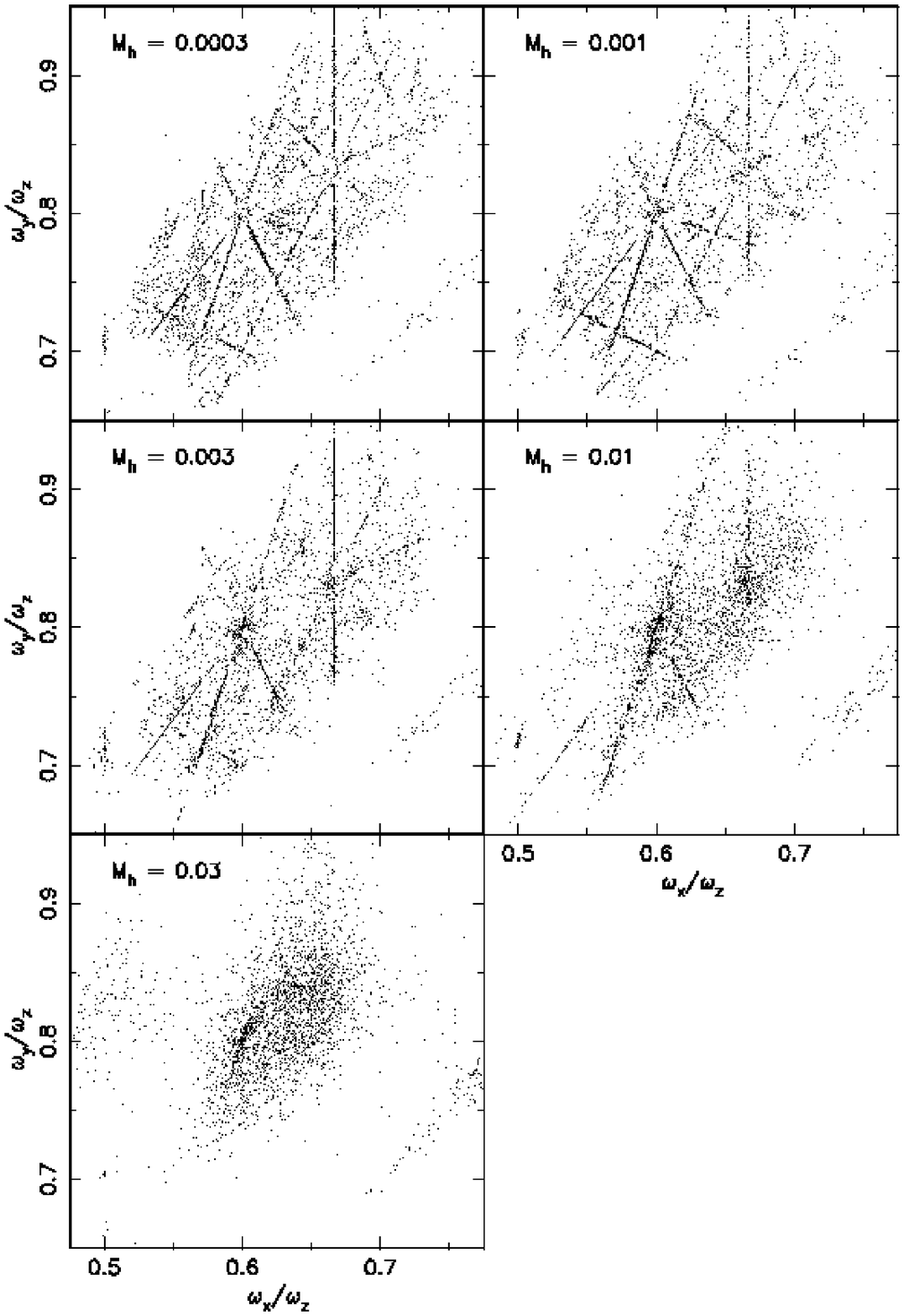}
\caption[]{
Frequency maps for the orbits of Figure \ref{fig_diff2}.
}
\label{fig_fm_2}
\end{figure*}

Figures \ref{fig_diff1} and \ref{fig_fm_1} show the diffusion rates and
frequency maps for ensembles of orbits with boxlike initial
conditions in a set of $\gamma$-models with five different values of
$\gamma$ between $0$ and $2$ and $c/a=0.5$.  The density cusps
in real elliptical galaxies are almost always found to have slopes
that lie in this range (\cite{mef95}; \cite{geb96}), with the steeper
cusps usually appearing in fainter galaxies.  
The diffusion rates have been normalized by dividing
the change in $\omega$ by the frequency of the long-axis orbit
in all of the models.  
Since the orbital integrations were carried out for a fixed number of
long-axis orbital periods, the diffusion rates plotted in Figure
\ref{fig_diff1} may be interpreted as the fractional change in $\omega$
that takes place over a fixed number ($50$) of orbital oscillations.

For $\gamma=0$, both the diffusion plots and the frequency maps reveal 
that the motion is largely regular, at least over the interval of $\sim 10^2$
oscillations for which the orbits were integrated.  The stochastic
orbits are mostly identified with the instability strip that connects
the $y$ and $z$ axes in initial condition space.  A number of orbits
that lie close in frequency space to a resonance between the three
degrees of freedom exhibit weaker stochasticity; the most prominent
resonance is the $(1,-2,1)$, which extends from the $y-z$ instability
strip through the unstable, $4:5:6$ and $3:4:5$ periodic orbits.
However the diffusion rates for most of these orbits are lower than
that in the $y-z$ instability strip by orders of magnitude.  As a
result, the majority of orbits outside of the $y-z$ instability strip
remain in a smooth grid in the frequency map.

As $\gamma$ is increased from $0$ to $1$, the area in initial
condition space associated with resonance layers becomes larger.
Almost all of the regular orbits in the $\gamma=1$ model can be
associated with a stable resonance, and a smaller number with stable
periodic orbits that lie at the intersection of two or more
resonances.  The regions between the resonance lines in the frequency
map now show almost no hint of a regular grid; in other words,
essentially none of the phase space in the $\gamma=1$ model can be
usefully identified with the integrable phase space of a St\"ackel
potential.  Nevertheless, the majority of the strongly stochastic
orbits remain confined to the $y-z$ instability strip; diffusion rates
in other parts of initial condition space are generally much lower.

The most dramatic change in the structure of phase space occurs as
$\gamma$ is increased from $1$ to $2$.  The $y-z$ instability strip
enlarges to include most of the equipotential surface; the diffusion
rates in this interconnected stochastic region become nearly constant,
implying that the stochastic motion is essentially free to explore the
entire phase-space region accessible to it.  Most of the stable
resonance zones disappear, leaving finally only the $(2,0,-1)$ and
$(3,-1,-1)$ resonances with appreciable numbers of associated regular
orbits.  In fact a large fraction of the regular orbits in the
$\gamma=2$ model appear to be closely associated with just three,
stable periodic orbits: the $1:2$ $x-z$ banana orbit 
(the $2,0,-1$ resonance); and the $4:5:7$ and $5:7:8$ orbits (both
associated with the $3,-1,-1$ resonance).  Essentially all of the
points on the frequency map that lie away from these two resonance
lines are stochastic.  However, a certain amount of structure remains
in stochastic phase space, since many of the fundamental frequencies
returned by the NAFF algorithm for the stochastic orbits are clustered
near one or more resonance lines.  In other words, over time 
invervals of
$\sim 10^2$ oscillations, many of the stochastic orbits continue to
behave to a certain extent like regular orbits, with more-or-less
well-defined frequencies.

Figures \ref{fig_diff2} and \ref{fig_fm_2} present the analogous 
results for ensembles of orbits in models with $\gamma=0.5$
and with five different values of $M_h$, the mass of a central singularity
in units of the total mass of the model.  
The progression from essentially regular to essentially chaotic 
motion is now more dramatic.  
Even for $M_h=0.0003$, the regular grid in the
frequency map is destroyed, and the motion is dominated by the
resonances.  As $M_h$ approaches $0.01$, the $y-z$ instability strip
enlarges to include most of the equipotential surface, and the
diffusion rates are uniformly high throughout most of stochastic
initial condition space.  The structure in the frequency map is almost
completely destroyed when $M_h=0.03$.  The only remaining regular
orbits are associated with the $x-z$ banana orbit and the $3:4:5$
periodic orbit, while the stochastic orbits are distributed almost
randomly around the frequency map.  In terms of the definitions
presented above, the phase space of the $M_h=0.03$ model (at least,
that part of phase space associated with boxlike orbits) appears to
be in the globally stochastic regime.  While the precise value
of $M_h$ corresponding to the transition to global stochasticity is
not clearly defined, Figures \ref{fig_diff2} and
\ref{fig_fm_2} suggest that it lies between $0.01$ and $0.03$ at 
this energy.

By comparison, even the steepest cusp ($\gamma=2$) appears to generate
no more stochasticity than a central singularity with $M_h\approx
0.003$.  In other words, in models without a central black hole, the
transition to global stochasticity appears to be just beginning as
$\gamma$ increases above $2$.  

The transition to global stochasticity seen in the boxlike
orbits is in sharp contrast to the behavior of the tubes.
Since the time-averaged angular momenta of tube orbits is nonzero,
tube orbits never go close enough to the center to experience the 
effects of a steeply-rising central force.
Consequently, one does not expect the structure of phase
space as seen in either diffusion rate plots or the frequency map
(Figure \ref{fig_map2}) of tube orbits to show much dependence 
on cusp slope. 
This was in fact found to be the case. 
The stochasticity in the tube orbits arises from the inherent
non-integrability of the model, and its character is determined 
by the overall shape of the potential, not by any peculiarity of 
the central density structure.

It was remarked above that regular orbits could be classified in 
terms of their degree of degeneracy, i.e. in terms of the number 
of independent resonance conditions 
$l\omega_1+m\omega_2+n\omega_3=0$ that define their associated 
phase-space regions.
Figures \ref{fig_diff1} and \ref{fig_diff2} suggest
that a systematic change takes place in the typical degree of 
degeneracy of the regular orbits as the strength of the perturbation 
is increased.
In weakly chaotic potentials (e.g. $\gamma=0$, $M_h=0$), 
most of the regular orbits lie in regions of zero-fold 
degeneracy, i.e. strongly non-resonant regions that have retained 
their integrable character in spite of the perturbation of the 
Hamiltonian away from integrable form.
As the perturbation is increased ($\gamma=0.5$), 
a larger fraction of the regular orbits lie in resonance zones, 
regions defined by a single resonance condition. 
When the perturbation is large ($\gamma=2$), the only remaining regular 
orbits lie in regions of two-fold degeneracy, i.e. regions 
associated with stable periodic orbits.
\section{Diffusion}
\subsection{Spectra of Diffusion Rates}

\begin{figure}
\vspace{7cm}
\includegraphics{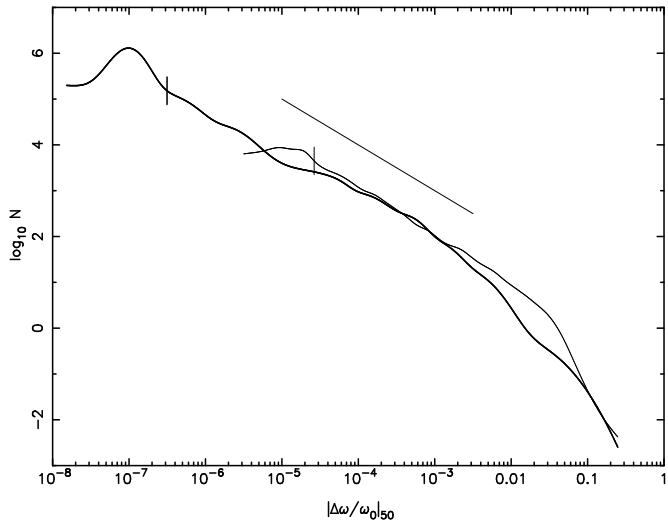}
\caption[]{
Spectra of diffusion rates for two ensembles of boxlike orbits at
shell 8 in the model with $\gamma=1$.
The thin curve was derived from $\sim 10^4$ orbits integrated for
100 orbital periods; the thick curve is from $\sim 10^3$ orbits
integrated for 400 periods.
Tick marks separate regular from stochastic orbits, as discussed
in the text.
The thin straight line has a logarithmic slope of $-1$.
}  \label{fig_hist0}
\end{figure}

While the distinction between regular and stochastic motion is
unambiguous in principle, stochastic orbits can mimic regular orbits
for very long times (\cite{con71}).  This is shown clearly in Figure
\ref{fig_hist0}, a plot of the distribution of diffusion rates for
boxlike orbits at shell 8 in the $\gamma=1$ model.  The thin curve
was derived from the same $\sim 10^4$ orbits whose frequency map is
displayed in Figure \ref{fig_fm_1}.  The solid curve is from a
smaller set of $\sim 10^3$ orbits integrated for a total of 400
long-axis orbital periods, four times the integration interval of the
larger ensemble.  (The accuracy of the numerical integration routine
was increased for the longer run.)  Since the precision with which
the NAFF algorithm computes the fundamental frequencies is a strong
function of the integration interval (\cite{las96}), we expect the
solid curve to extend to smaller values of $\Delta\omega$ than the
thin curve, assuming of course that very weakly chaotic orbits exist
in this potential.  The abscissa, $|\Delta\omega/\omega_0|_{50}$, is
the change in fundamental frequencies over an interval of 50
long-axis orbital periods, normalized to $\omega_0$, the frequency of 
the long-axis orbit at shell 8.
In the longer run, the measured changes in $\omega$ were scaled 
to the shorter interval following the procedure described in
\S6.

The spectrum of diffusion rates is well described as a power law in
$\Delta\omega$ with logarithmic slope close to $-1$.  In the run
with the longer integration interval (solid curve), this power law
distribution extends over at least six decades in $\Delta\omega$, a
remarkable range.  Evidently, an appreciable fraction of the
stochastic orbits in this model exhibit motion that is very nearly,
though not quite, regular.  
This fact makes it difficult to decide
which orbits are regular -- there is no well-defined break in the
spectrum at low $\Delta\omega$.  Presumably, many of
the orbits with $\Delta\omega$ below the minimum value detectable by
the NAFF routine are regular.  Based on the accuracy tests described
above, we estimated, for the shorter run, the smallest significant
value of $\Delta\omega$.  This value, indicated by the thin vertical
tick mark, implies that $\sim 22\%$ of the orbits in the shorter run
are effectively regular.  A similar tick mark has been placed on the
solid curve at the 22\% point.  Happily, on both curves, the tick
mark falls roughly at the point where the distribution turns upward:
sharply in the case of the longer run, more weakly in the case of the
shorter run.  Plots of the starting points on the equipotential
surface of orbits to the left of the tick marks also show a very
similar distribution for the two ensembles.  We conclude that roughly
$1/5$ of the orbits in each ensemble are regular, or so weakly
stochastic that we can not distinguish them from regular orbits.

One might reasonably ask whether there are {\it any} regular orbits
in these ensembles.  
Perhaps the upturn at small $\Delta\omega$ in
Figure \ref{fig_hist0} reflects a population of orbits with extremely
low but nonzero diffusion rates.  
We consider such an interpretation to be unlikely.  
Certain of the periodic orbits in this initial condition
space (e.g. the $x-z$ banana) are known to be stable to small
perturbations (\cite{frm97}) and must have families of associated
regular orbits.  
The orbits with the lowest diffusion rates do in
fact have initial conditions lying close to these stable periodic
orbits.  
A number of experiments reassured us that the
upturn at small $\Delta\omega$ in the distribution of diffusion rates
always occurs at roughly the same percentile point in a given
initial-condition space, regardless of (modest) changes in the
integration interval or accuracy of the numerical integrator.  
If there were no bona-fide regular orbits, we would expect to see the
fraction of orbits with significantly nonzero $\Delta\omega$'s
increase monotonically with increasing accuracy of the numerical
routines.

In any case, we clearly see a smaller population of regular box
orbits here than was seen in many earlier studies.  For instance,
Merritt \& Fridman (1996) used Liapunov exponents to study orbits in
the same $\gamma=1$ model investigated here.  In their
ensemble of 192 boxlike orbits from shell 8, they estimated that
$\sim 40\%$ were regular (their Figure 7a), roughly twice the
fraction of regular orbits found here.  This difference is not
surprising given the greater sensitivity of the NAFF algorithm to
weak stochastiticy.  Figure \ref{fig_hist0} implies that Merritt \&
Fridman (1996) were unable to detect stochasticity in orbits with
$\Delta\omega/\omega_0$ less than about $3\times 10^{-4}$.  As we
show below, such small amounts of diffusion correspond to very nearly
regular behavior.

\begin{figure}
\vspace{12cm}
\includegraphics{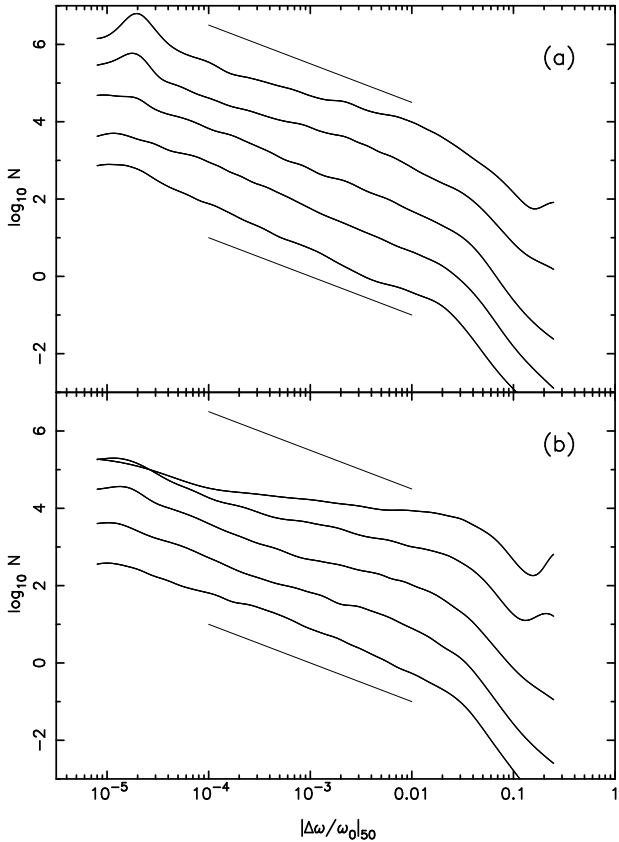}
\caption[]{
Spectra of diffusion rates for the ten ensembles of boxlike orbits.
(a) $M_h=0$; $\gamma=0,0.5,1,1.5,2$, increasing upward.
(b) $\gamma=0.5$; $M_h=0.0003,0.001,0.003,0.01,0.03$, increasing upward.
The thin straight lines have power-law slopes of $-1$. The curves are
offset vertically with respect to each other by one unit in the ordinate.
}  \label{fig_hist1}
\end{figure}

The existence of a population of stochastic orbits with very low
diffusion rates is consistent with a number of studies of motion in
``decomposable'' phase spaces, i.e. systems containing both regular
and stochastic trajectories (e.g. \cite{kar83}).  In decomposable systems,
stochastic orbits are hampered in their diffusion by invariant tori;
when a large fraction of phase space is associated with such tori,
many of the stochastic orbits exhibit quasi-regular behavior over
long periods of time.

The approximately $1/\Delta\omega$ distribution of diffusion rates in
Figure \ref{fig_hist0} can not extend to very large or very small
$\Delta\omega$'s without producing a logarithmic divergence in the
total number of orbits.  In the spectrum derived from the shorter
run, we in fact observe a dropoff for
$|\Delta\omega/\omega_0|_{50}\gap 0.05$.  This is expected, since the
fundamental frequencies associated with orbits at a given energy are
restricted to a range of values near the oscillation frequencies
along the major axes, which means that $\Delta\omega/\omega_0$ can
never exceed unity and will generally be much smaller.  There must
also be a falloff for small diffusion rates, but this apparently
occurs at such small values of $\Delta\omega$ that it is obscured by
the contribution from regular orbits.

Figure \ref{fig_hist1} shows the distribution of diffusion rates for
the full set of 10 ensembles whose frequency maps are displayed in
Figures \ref{fig_fm_1} and \ref{fig_fm_2}.  As the degree of central
mass concentration is increased, the spectra become
shallower, i.e. a larger fraction of the orbits are strongly
diffusing.  The logarithmic slope decreases from $\sim -1$ for
$\gamma\lap 1$ to $\sim -0.85$ for $\gamma=1.5$ and $\sim -0.75$ for
$\gamma=2$.  There is a corresponding decrease in the fraction of
regular orbits: from $\sim 32\%$ for $\gamma=0$ to $\sim 18\%$ for
$\gamma=2$.  In the models with a central point mass, the spectra are
better described as two power laws, with the shallower power law
characterizing the larger diffusion rates.  For $M_h=0.001$, the
shallower power law has a logarithmic slope of $\sim -0.85$ ($\sim
15\%$ regular), increasing to $\sim -0.56$ for $M_h=0.01$ ($\sim
12\%$) and $\sim -0.32$ for $M_h=0.03$ ($\sim 5\%$).  We stress that
our orbits are not selected in a uniform way from the energy
hypersurface and the distribution of diffusion rates that we find
need not be characteristic of phase space as a whole.  Nevertheless
the lesson of these plots seems clear: as one increases the central
mass concentration of a triaxial model, the fraction of regular
orbits drops and the typical diffusion rate goes up.

\subsection{The Character of Diffusion in Frequency Space}

\begin{figure}
\vspace{8cm}
\includegraphics{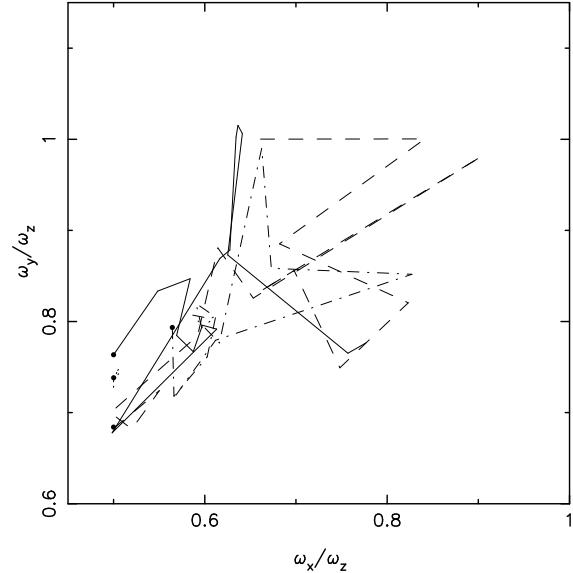}
\caption[]{
Diffusion in frequency space of four stochastic orbits
in the model with $\gamma = 2$ and $M_h=0$.
All the orbits have similar values of
$|\Delta \omega/\omega_0|_{50}$.
Solid dots mark the starting points; line segments connect
positions in the frequency map obtained at intervals of 50 orbital
periods, for a total of 800 periods.
}  \label{fig_rwk1}
\end{figure}

As a stochastic orbit evolves, it visits different parts of
frequency space, eventually moving far away from its initial 
location.
In Figure \ref{fig_rwk1} we plot the trajectory in frequency space 
of four orbits in the $\gamma=2$ model.
The solid dot in each curve marks the initial location of the orbit;
successive positions in frequency space, at intervals of 50 orbital 
periods, are joined by line segments. 
For three of the orbits, the motion in frequency space is reminiscent 
of a random walk, in the sense that successive displacements are not 
strongly correlated. 
The fourth orbit appears to diffuse far less than the other three. 
A comparison with the frequency map of Figure \ref{fig_fm_1} shows 
that this orbit is located close to the $x-z$ banana and is 
evidently trapped in the associated resonance layer. 

\begin{figure}
\vspace{13cm}
\includegraphics{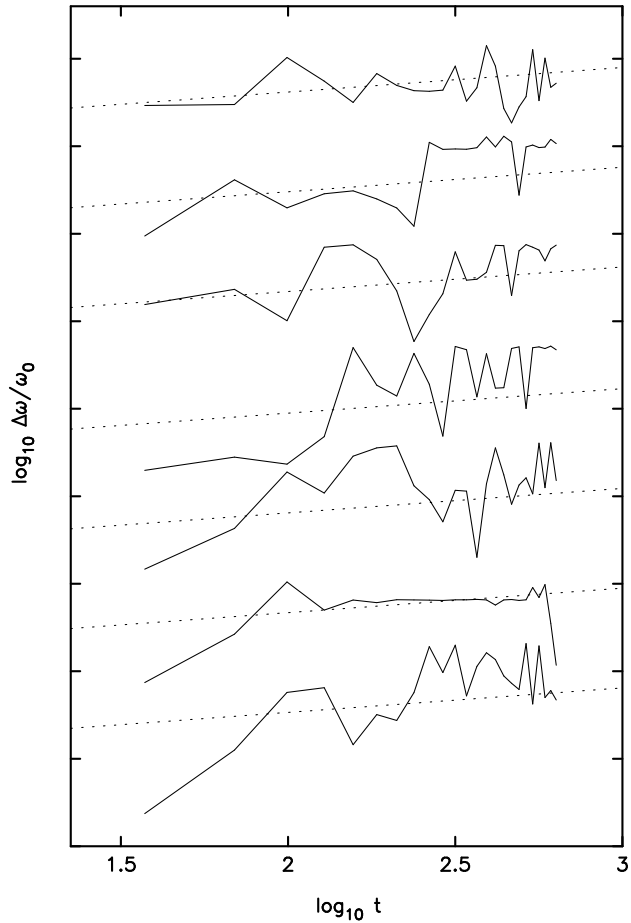}
\caption[]{
Cumulative change in fundamental frequencies versus time 
for seven stochastic orbits.
The straight lines have logarithmic slopes of $0.5$.
}  \label{fig_rwk2}
\end{figure}

Assuming a random walk in frequency space, the distance that an 
orbit moves from its starting point should increase
approximately as $t^{1/2}$, with $t$ the elapsed time.  
In Figure \ref{fig_rwk2} we show the time evolution of 
$\Delta\omega$ (as defined above) for 7 different 
orbits from the model with $\gamma = 0.5$ and $M_h = 0.03$.
These 7 orbits are representative of a larger sample of 100 that
were randomly selected from the large ensemble of $10^4$ orbits
discussed previously, on the basis that their diffusion
over 50 orbital periods satisfied $1\times 10^{-4} \leq
|{\Delta \omega}|/\omega_0 \leq 3\times 10^{-4}$.
The dotted line has a logarithmic slope of $1/2$.
While there are large excursions in the value of $\Delta\omega$ 
for each of the orbits, there is a reasonably clear trend for 
$\Delta\omega$ to increase monotonically with time with an
approximately $t^{1/2}$ dependence. 
For orbits with intially larger values of $\Delta \omega/\omega_0$,
the diffusion was found to saturate rapidly, as expected given 
that frequency space at a given energy is bounded.

While it is beyond the scope of this paper to investigate the 
character of stochastic diffusion in more detail, a limited 
number of experiments convinced us that the diffusion of boxlike 
orbits in moderately stochastic potentials -- corresponding to 
models with $\gamma\gap 1$ or $M_h\gap 0.003$ -- can often be 
approximated as a random walk.
We will make use of this result in the following section.

\subsection{The Consequences of Diffusion for the Shapes of Orbits}

\begin{figure*}
\vspace{21cm}
\includegraphics{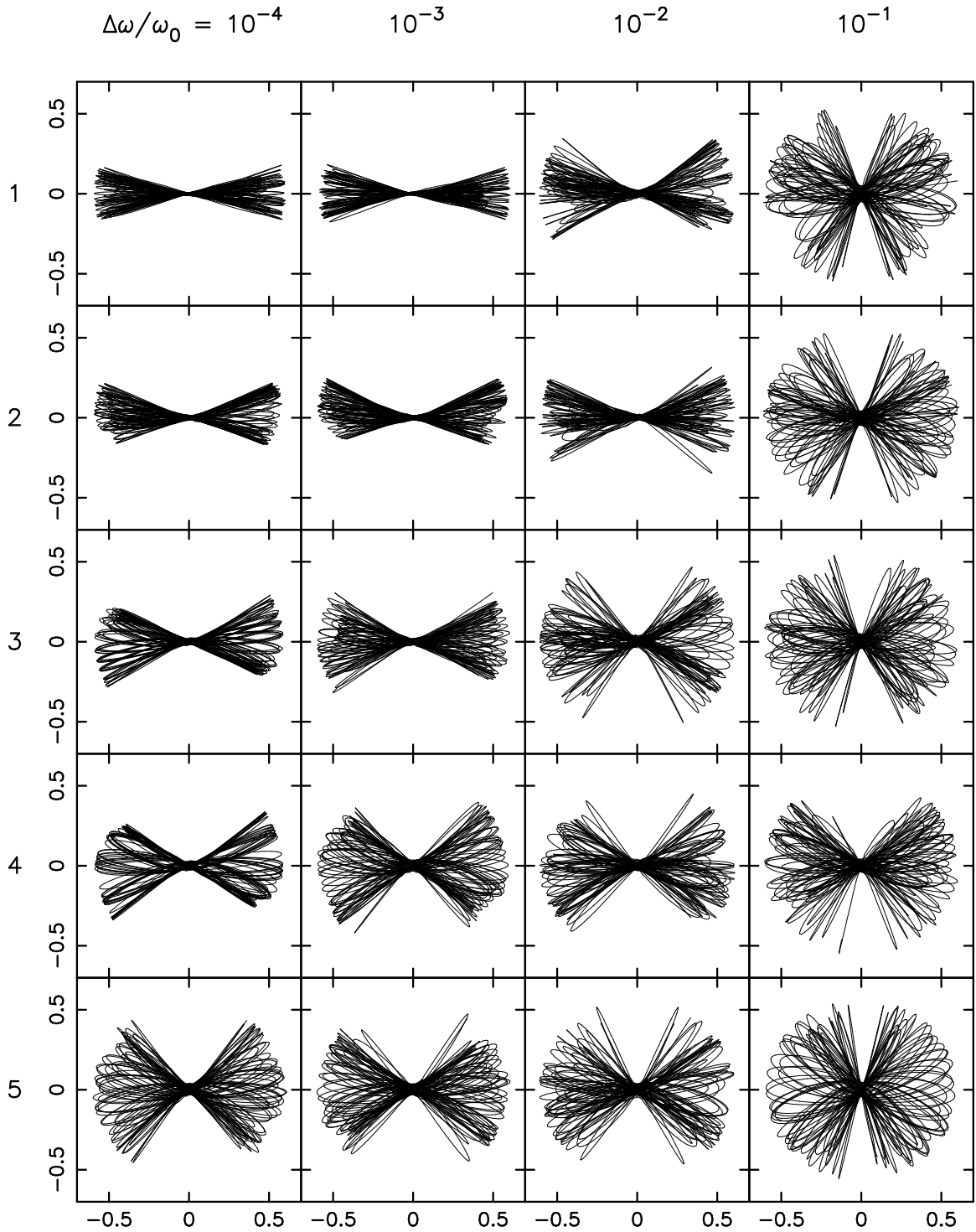}
\caption[]{
Boxlike orbits from shell 8 of the model with $\gamma=2$, integrated
for 50 orbital periods.
The abscissa and ordinate are the $x$ and $y$ axes in each plot.
Initial conditions were drawn from one of five small patches on the
equipotential surface, as described in the text and labelled at the left.
In each column, orbits were chosen to have similar values of 
$\Delta\omega/\omega_0$, as indicated at the top.
}
\label{fig_orbits}
\end{figure*}

Orbits that diffuse very slowly through frequency space are
effectively regular: they maintain a well-defined shape for
long periods of time.
Stochastic orbits are significant from the standpoint of galaxy
evolution only if the configuration-space volume through which they
move changes substantially over the time period of interest.  
We therefore need to ask how large a value of $\Delta\omega/\omega_0$
corresponds to a significant change in the configuration-space volume
filled by an orbit.  
To answer this question, we selected sets of orbits from the 
$\gamma=2$ ensemble having a wide range of $\Delta\omega$'s, but 
lying within small patches on the equipotential surface.  
Orbits which originate in a small region would be
expected to have similar shapes if not for the fact that they were
stochastic. 
By comparing orbits within such a set, we can estimate how much 
change in configuration space density is implied by
a given value of $\Delta \omega$.  

The orbits were integrated for 50 periods and their configuration-space 
trajectories recorded.  
Figure \ref{fig_orbits} presents a selection of orbits
taken from five different patches on the equipotential surface.
Region 1 lies near the starting point of the $x$-axis orbit, while
region 5 lies above the $x-y$ plane roughly halfway between the $x$
and $y$ axes.  In a fully integrable potential, these starting points
would all be associated with narrow box orbits, orbits that are
elongated in the same sense as the figure.  Narrow box orbits are often
found to be heavily populated in self-consistent triaxial models
(\cite{sch79}) and any significant evolution in their shapes would be
expected to have important consequences for a triaxial galaxy.

Figure \ref{fig_orbits} shows that the degree of orbital evolution
is crudely predictable given the change in fundamental frequencies.  
For $\Delta\omega/\omega_0 \approx 10^{-4}$, the orbits
are almost indistinguishable from regular orbits, accurately
maintaining their elongated shapes.  
For $\Delta\omega/\omega_0\approx 10^{-3}$, the orbits are noticeably 
irregular, and for $\Delta\omega/\omega_0 \approx 10^{-1}$ the orbits 
have almost entirely lost their elongated characters.  
From these experiments, we
estimate that a fractional change in fundamental frequencies of order
$\Delta\omega/\omega_0\approx 0.03$ corresponds to a significant
change in the configuration-space shape of an orbit.  If a large
fraction of the boxlike orbits in a triaxial model experienced
changes of this order over some specified length of time, we might
expect the model to evolve strongly in shape over the same interval.

\begin{figure}
\vspace{12cm}
\includegraphics{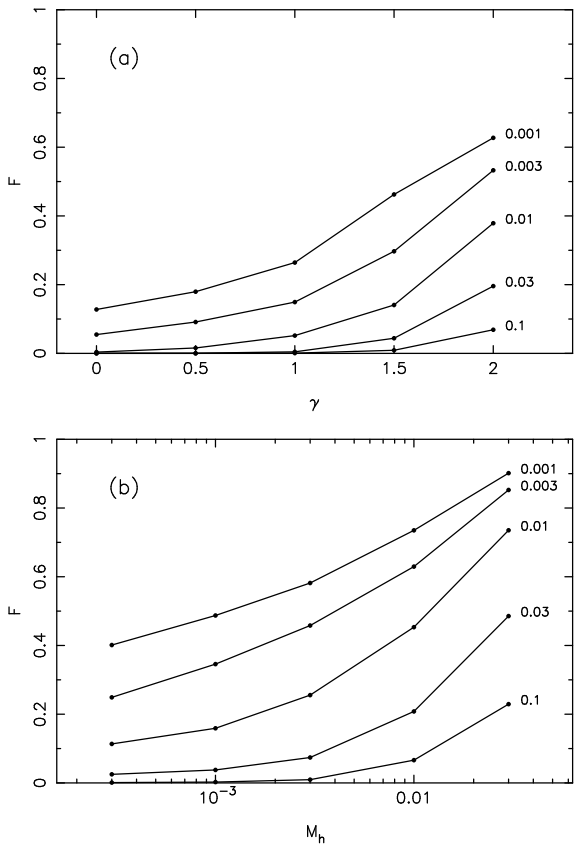}
\caption[]{
Fraction of boxlike orbits at shell 8 with $|\Delta\omega/\omega_0|_{50}$ 
exceeding some fiducial value, as indicated on the right.
(a) $M_h=0$; (b) $\gamma=0.5$.
}  \label{fig_stat1}
\end{figure}

Figure \ref{fig_stat1} shows how the fraction $F$ of orbits with
$|\Delta\omega/\omega_0|_{50}$ exceeding some fiducial value varies
with $\gamma$ and $M_h$ in the ensembles of Figure \ref{fig_fm_1}.
In models without a black hole, a modest fraction of the boxlike
orbits, $F\approx 20\%$, have evolved strongly after 50 orbital
periods when the cusp is steep, $\gamma\approx 2$.  In the models
with a black hole, the fraction of boxlike orbits with
$|\Delta\omega/\omega_0|_{50}>0.03$ reaches $\sim50\%$ for
$M_h=0.03$, and exceeds 20\% for all $M_h\gap 0.01$.  Fully 75\% of
the boxlike orbits -- i.e., essentially all of the stochastic
boxlike orbits -- have $|\Delta\omega/\omega_0|_{50}>0.01$ in the
$M_h=0.03$ model, as expected if this initial-condition space is in
the globally-stochastic regime.

We discuss in more detail below how one might translate these numbers
into rough estimates of the time over which a triaxial galaxy
would evolve in shape.  However, a few conclusions follow
unambiguously from Figure \ref{fig_stat1}.  Even a relatively modest
black hole, $M_h\approx 0.003$, induces about as much stochastic
diffusion in the boxlike orbits as does a steep central cusp,
$\gamma=2$.  Since the average ratio of black hole mass to galaxy
mass is believed to be about 0.005 (\cite{kor95}), Figure
\ref{fig_stat1} suggests that the chaotic evolution in most
early-type galaxies will be driven more by the central black hole
than by the stellar cusp.

The greater effectiveness of black holes compared with cusps at 
inducing chaos is not surprising.
Miralda-Escud\'e \& Schwarzschild (1989) point out that the 
gravitational force from even a steep density cusp,
$\rho\propto r^{-2}$, fails to produce the large-angle deflections 
that result from close passage to a point mass.

\subsection{Dependence of Diffusion Rates on other Parameters}

Up till now, all of our results were derived from orbits at shell 8,
whose apocenters lie just inside the half-mass radius of the model.
Furthermore we have only discussed a single, relatively elongated
mass model, with $c/a=0.5$.  
Most elliptical galaxies are less elongated than this (\cite{trm95}).  
Here we extend our results
to a range of different energies in models with two additional
shapes: $c/a=0.6$ and $0.8$.  
We continue to assume ``maximal triaxiality.''  
In each model, we carried
out a frequency analysis of $\sim500$ boxlike orbits at each of five
energies, corresponding to shells $3, 6, 9, 12$ and $15$.  As before,
we considered five different values of the cusp slope $\gamma$ and
five different values of the black hole mass $M_h$ (the latter in
models with $\gamma=0.5$).  Orbits were integrated for 50 periods of
the long-axis orbit at the energy corresponding to their shell.

\begin{figure}
\vspace{12cm}
\includegraphics{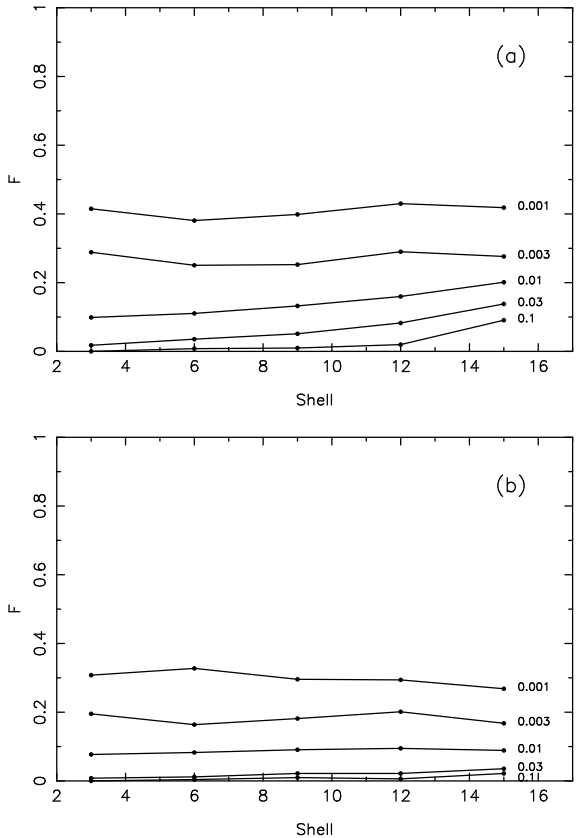}
\caption[]{
Fraction of boxlike orbits with $|\Delta\omega/\omega_0|_{50}$ 
exceeding some fiducial value, as a function of shell number,
for $\gamma=1.5$ and $M_h=0$.
(a) $c/a=0.6$; (b) $c/a=0.8$.
}  \label{fig_stat_g}
\end{figure}

In the models without a central black hole, the spectrum of diffusion
rates was found to vary only modestly from shell to shell.  In other
words, boxlike orbits experience roughly the same degree of
evolution, after a fixed number orbital periods, at all energies.
(Since orbital periods are shorter near the center of every model,
this result implies a {\it greater} degree of evolution at low
energies after a fixed interval of time.)  The greatest dependence on
energy was seen in models with a weak cusp, $\gamma\approx 0$; these
models have nearly-harmonic cores and the stochasticity becomes less
important at low energies.  In models with $\gamma\gap 1$, almost no
variation with energy is seen (Figure \ref{fig_stat_g}). 

\begin{figure}
\vspace{12cm}
\includegraphics{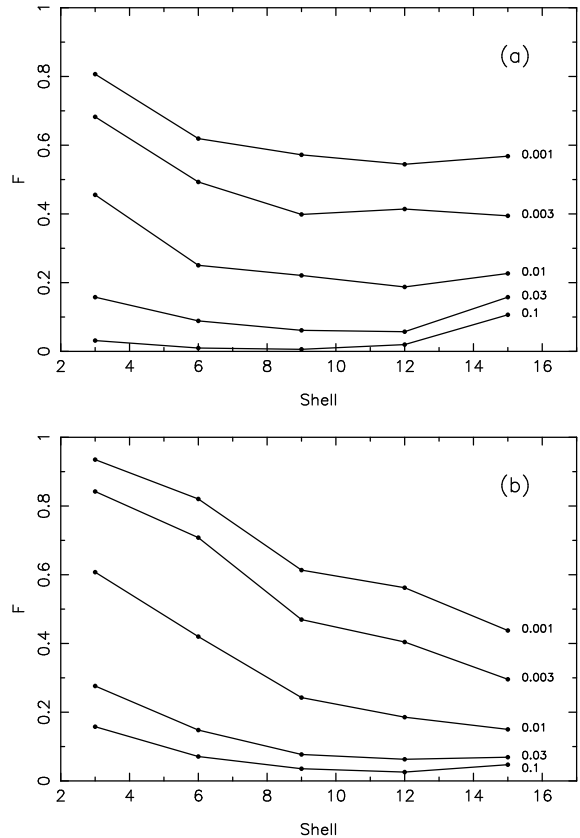}
\caption[]{
Fraction of boxlike orbits with $|\Delta\omega/\omega_0|_{50}$ 
exceeding some fiducial value, as a function of shell number,
for $M_h=0.003$ and $\gamma=0.5$.
(a) $c/a=0.6$; (b) $c/a=0.8$.
}  \label{fig_stat_b}
\end{figure}

In models with a central black hole, there is a stronger dependence
of diffusion rates on energy (Figure \ref{fig_stat_b}).  At low
energies, an increasingly large fraction of the boxlike orbits
exhibit strong diffusion over a fixed number of orbital periods.
This is reasonable, since at low energies the black hole contains a
large fraction of the interior mass.

Figures \ref{fig_stat_g} and \ref{fig_stat_b} also indicate a modest
dependence of diffusion rates on galaxy elongation
$c/a$.  However, this dependence is likely to be swamped by the fact
that the boxlike orbits occupy a rapidly decreasing fraction of
phase space in more axisymmetric potentials.

One might expect the critical value of $M_h$ for transition to 
global stochasticity (\S 4) to depend on $c/a$ and on energy.
If we define a globally-stochastic initial-condition space as one in which 
5\% or less of the boxlike orbits are regular -- roughly the 
fraction at shell 8 in the model with $M_h=0.03$ (Figure 
\ref{fig_diff2}) -- we find that the transition takes place 
at shell 3 when $M_h\approx 0.005$, at shell 6 when $M_h\approx 0.01$,
at shell 9 when $M_h\approx0.02$ and at shell 12 when $M_h\approx 0.03$
in the model with $c/a=0.6$.
When $c/a=0.8$, we find $M_h\approx 0.002$ at shell 3, $\sim 
0.004$ at shell 6, $\sim 0.01$ at shell 9 and $\sim 0.02$ at 
shell 12.

We note that diffusion rates of stochastic orbits in real 
galaxies might depend on additional factors not considered here, 
such as the discreteness of the potential or its dependence on 
time.
For instance, Habib, Kandrup \& Mahon (1997) have argued that even 
very small amounts of additive noise can greatly enhance the diffusion 
rate in 2 DOF systems.
We hope to address this issue in a subsequent paper.

\section{Evolution}

Diffusion of stochastic orbits is an irreversible process.
Because chaotic motion is essentially random over long time 
intervals, the probability of finding a single star anywhere in 
stochastic phase space tends toward a constant value at all accessible 
points; in other words, the density of an ensemble of stars in 
stochastic phase space evolves toward a constant, coarse-grained 
value.
Something similar to this takes place in regular phase space as 
stars on nearby orbits gradually move out of phase.
But chaotic mixing is more efficient than phase mixing 
because the region accessible to a stochastic orbit is much larger 
than the single torus to which a regular orbit is confined, and 
because the divergence between adjacent stochastic trajectories grows 
exponentially at early times (\cite{kam94}; \cite{mev96}).
In a fully mixed galaxy, there is only one ``orbit'' or 
``invariant density'' at each energy in stochastic phase space,
and this orbit has a configuration-space shape that is
poorly suited to reproducing the shape of the galaxy (\cite{mef96}).
If diffusion times for a significant fraction of the 
orbits in a triaxial galaxy are short compared to the galaxy's 
lifetime, the galaxy would be expected to evolve toward more 
spherical or axisymmetric shapes as the stochastic parts of phase space 
approach a fully mixed state.

We wish to estimate the approximate rate of this evolution.
In particular, we want to know how the rate of
chaos-driven evolution varies with galaxy parameters.
From a dynamical point of view, early-type galaxies comprise a 
two-parameter sequence, the so-called Fundamental Plane 
(\cite{pah95}).
But to a first approximation the structural parameters of 
early-type galaxies can be expressed as a function of just one 
variable, the total luminosity.
Bright elliptical galaxies and bulges have lower average densities, 
larger characteristic radii and shallower central density cusps than 
faint ellipticals.
These trends imply that chaotic mixing should be most effective
in faint (triaxial) galaxies, for at least two reasons.
First, crossing times are shortest in faint ellipticals, and 
stochastic diffusion rates scale approximately with orbital 
frequency.
Second, faint ellipticals have higher central concentrations, 
implying a greater degree of stochastic evolution for boxlike orbits
after a given number of orbital periods.
Bright ellipticals might also be dynamically younger than
faint ellipticals in the sense of having formed more recently
via mergers.
We will argue below that the greater effectiveness of chaotic 
mixing in fainter ellipticals may be responsible for many of 
the systematic differences between them and bright ellipticals, 
including their lower degree of triaxiality and their less boxy
isophotal shapes.

In this section, we use the diffusion-rate calculations of \S5 
to estimate how great a degree of orbital evolution would be 
expected after a Hubble time in real elliptical galaxies.
We begin (\S6.1) by using observational data to estimate the average 
dependence of galaxy crossing time on luminosity, which allows us 
to determine how the ``dynamical ages'' of ellipticals vary with 
their absolute magnitude.
Since our calculations in \S5 yielded spectra of diffusion rates 
after a fixed interval of $50$ orbital periods, we need to 
find a way to scale those spectra to different elapsed times.
We do this (\S6.2) by assuming that diffusion in frequency space can be 
approximated as a random walk.
We can then calculate the fraction of boxlike orbits that 
would have experienced substantial evolution over a galaxy 
lifetime as a function of galaxy luminosity.

Evolution of individual orbits is significant only if phase space 
is populated in a non-uniform way to start with; in a fully mixed 
galaxy, stochastic diffusion would leave the phase space density 
unchanged, i.e. constant in each disjoint region.
Our assumption throughout this section is that galaxies form in a 
state that is not fully mixed, and hence that diffusion of 
stochastic orbits can lead to a change in a galaxy's shape.
We clearly do not have access here to the tools that would be 
required to accurately calculate the rate of such change; at a minimum, we 
would need to know the orbital population of a self-consistent 
model before we could convert our orbital diffusion rates into 
rates of change of the self-consistent shape.
Nevertheless we can make some tentative statements about the 
expected rate of evolution by comparing our results to 
a recent set of $N$-body simulations of triaxial galaxies with 
central black holes (\S6.3).

\subsection{Dependence of Elliptical Galaxy Structural Parameters on Luminosity}

\begin{figure}
\vspace{7cm}
\includegraphics{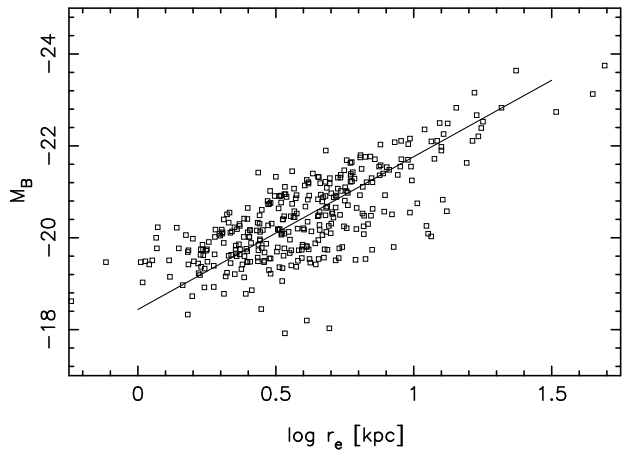}
\caption[]{
Relation between absolute blue magnitude and effective radius for
a sample of 337 nearby E and S0 galaxies, from the data of
J\o rgensen et al. (1995, 1996).
The line is a least-squares fit, as described in the text.
}  \label{fig_inger}
\end{figure}

The observed variation of effective radius $r_e$ with blue
absolute magnitude $M_B$
is shown in Figure \ref{fig_inger} for a sample of 337 early-type 
galaxies, from the data of J\o rgensen et al. (1995, 1996).
The fitted line has
\begin{equation}
M_B = -18.440 - 3.327\log r_e;
\label{jorgen1}
\end{equation}
the slope in Equation (\ref{jorgen1}) is the geometric mean of 
the slopes obtained by least-squares fits to $M_B$ vs $r_e$ and 
$r_e$ vs $M_B$.
The mean relation can be written
\begin{equation}
r_e \approx 12.50\left({L_B\over 10^{11}\Lsolar}\right)^{0.751} 
{\rm kpc}
\label{jorgen2}
\end{equation}
with $L_B$ and $\Lsolar$ the blue luminosities of the galaxy 
and the sun, respectively.
In the $\gamma$-models, the effective radius remains an approximately
fixed multiple of the half-mass radius as $\gamma$ is varied, 
$r_e\approx 0.75 r_{1/2}$ (\cite{deh93}), which allows us to 
replace $r_e$ with $r_{1/2}$ in equation (\ref{jorgen2}).

For the dependence of mass-to-light ratio on luminosity, 
we take
\begin{equation}
{M\over L_B} = 15.3\left({L_B\over 
10^{11}\Lsolar}\right)^{0.25}
\label{moverl1}
\end{equation}
in solar units (\cite{fab87}), or
\begin{equation}
{M\over\Msolar} = 1.53\times 10^{12}\left({L_B\over 
10^{11}\Lsolar}\right)^{1.25}.
\label{moverl2}
\end{equation}
The $M/L\sim L^{1/4}$ dependence of equation (\ref{moverl1}) could
also have been arrived at by combining equation (\ref{jorgen2}) with 
the virial theorem and the Faber-Jackson law.

If we define $T_{1\over 2}$ as the period of a circular orbit in Dehnen's 
spherical model at the radius containing one-half of the 
total mass, equations (\ref{jorgen2}) and (\ref{moverl2}) give
\begin{equation}
T_{1/2} \approx 2.30\times 10^{8}\left({L_B\over 
10^{11}\Lsolar}\right)^{0.50} {\rm yr},
\label{dynam}
\end{equation}
independent of $\gamma$.
This relation implies a modest dependence of galaxy crossing time on 
absolute magnitude.
For $M_B=-21$, $T_{1/2}\approx 1.4\times 10^8$ yr, or $\sim 35$ orbital
periods after a lifetime of $5\times 10^9$ yr.
For $M_B=-18$, $T_{1/2}\approx 3.47\times 10^7$ yr, for a dynamical
age of $\sim 150$ periods.
Finally, we can express $T_{1/2}$ in terms of the period of the axial orbit 
at shell 8 in our models, the unit of time in which results were
presented above.

\subsection{Scaling of $\Delta\omega$ to Different Time Intervals}

The integration interval on which Figures 
\ref{fig_hist1} and \ref{fig_stat1} are based corresponds roughly 
to the lifetime, at the effective radius, of a bright elliptical 
galaxy.
The orbital periods in fainter ellipticals are generally shorter.
In order to draw conclusions about the degree of evolution to be 
expected in galaxies with a range of properties, we need to 
scale our results on stochastic diffusion to different intervals 
of elapsed time.

The experiments in \S5.2 suggest that diffusion in frequency space 
can be reasonably approximated as a random walk over limited periods 
of time.
Define $x=|\Delta\omega/\omega_0|_{50}$, the fractional change in 
fundamental frequencies of a single orbit as computed over a time 
interval of $t_{50}$, equal to 50 orbital periods at shell 8.
The spectra of diffusion rates plotted in Figure \ref{fig_hist1} are 
plots of $N(x)$.
Assuming a random walk, the displacement of an 
orbit from its initial position in frequency space increases 
approximately as $t^{1/2}$.
After some time interval $t>t_{50}$, one expects to find
\begin{equation}
{\Delta\omega\over\omega_0} \approx x\left({t\over t_{50}}\right)^{1/2}.
\label{defdw}
\end{equation}
Equation (\ref{defdw}) -- which is only intended to be correct in a 
statistical sense -- allows us to estimate the change in 
fundamental frequencies $\Delta\omega$ after an elapsed time $t$ 
given a measured $|\Delta\omega|_{50}=x\omega_0$.
For $\Delta\omega/\omega_0\approx 1$, equation (\ref{defdw}) must 
break down, as discussed above; indeed, the NAFF estimate of the 
fundamental frequencies is only meaningful if the change in the 
$\omega$'s over the integration interval is fractionally small 
(\cite{las93}).

Let $\Delta\omega_c$ be the change in fundamental frequencies 
corresponding to a substantial evolution in the 
configuration-space shape of an orbit.
It was argued above (see the discussion accompanying Figure \ref{fig_orbits}) 
that $\Delta\omega_c\approx 0.03 \omega_0$.
After an elapsed time $t$, the fraction $F$ of orbits with 
$\Delta\omega>\Delta\omega_c$ is the fraction from the original 
spectrum $N(x)$ which satsify
\begin{equation}
x > {\Delta\omega_c\over\omega_0}\left({t\over 
t_{50}}\right)^{-1/2}.
\label{xgt}
\end{equation}
Our few experiments using NAFF with different integration intervals 
produced results that were consistent with equation (\ref{xgt}), 
though we can hardly claim to have confirmed this equation in any 
very general way. 
Equation (\ref{xgt}) -- which was used to scale the two spectra in Figure 
\ref{fig_hist0} -- allows us to compute the spectrum of 
$\Delta\omega$'s after any elapsed time given a measured 
spectrum $N(x)$.

\begin{figure}
\vspace{12cm}
\includegraphics{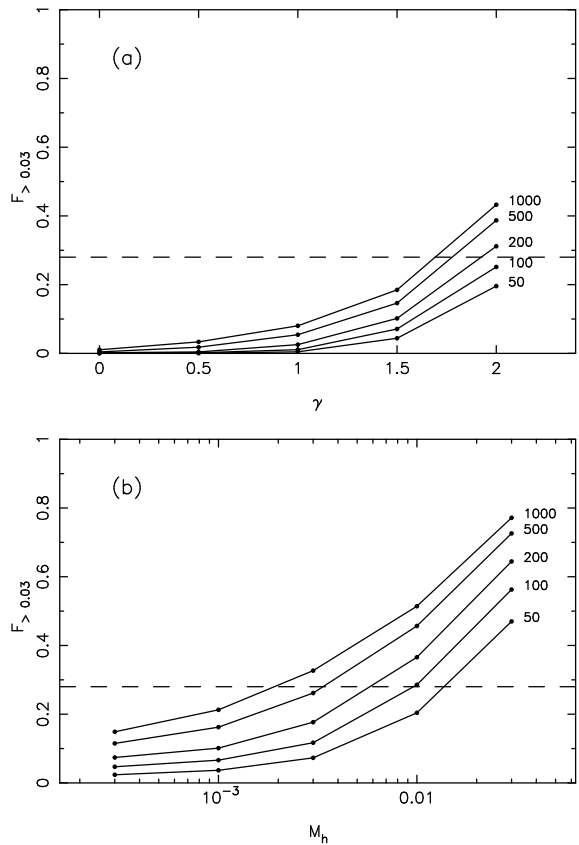}
\caption[]{
Fraction of boxlike orbits at shell 8 with $\Delta\omega/\omega_0$ 
exceeding 0.03 after some elapsed number of orbital periods at shell 8, 
as indicated on the right.
The horizontal dashed line at $F=0.28$ is discussed in the text.
}  \label{fig_stat2}
\end{figure}

In the case of a $1/\Delta\omega$ distribution of diffusion 
rates between some limiting values $x_1$ and $x_2$, 
equation (\ref{xgt}) implies
\begin{equation}
F(t) \approx 1 - \log\left[{\Delta\omega_c\over\omega_0} 
{1\over x_1}\left({t_{50}\over t}\right)^{1/2}\right]
/\log\left({x_2\over x_1}\right),
\label{frac}
\end{equation}
i.e. the fraction of strongly-evolved orbits increases 
logarithmically with time until 
$t\approx(\Delta\omega_c/\omega_0x_1)^2 t_{50}$, at which point
the phase-space distribution is ``fully mixed.''
This predicted behavior is consistent with what has been 
observed in a few earlier studies.
For instance, Merritt \& Valluri (1996) computed Liapunov 
exponents for boxlike orbits in a family of triaxial models and 
found that the fraction of orbits judged ``stochastic'' -- i.e, 
orbits having significantly nonzero Liapunov exponents -- 
gradually continued to increase as the integration interval was lengthened.
At the half-mass energy in their model with $m_0=0.001$, 
which has a nearly $r^{-2}$ density cusp,
Merritt \& Valluri found that $\sim 65\%$ of the orbits were 
``stochastic'' over an integration interval of $10^2$ periods; 
$\sim 72\%$ over $10^3$ periods; and $\sim 79\%$ over $10^4$ 
periods.
Thus, the number of orbits behaving chaotically increased by roughly
a constant amount following each factor of ten increase in the
integration time, roughly as expected for a $\sim 1/\Delta\omega$
distribution of diffusion rates.

In Figure \ref{fig_stat2}, equation (\ref{xgt}) has been used to 
estimate the fraction of strongly-evolved orbits (orbits with 
$\Delta\omega/\omega_0>0.03$) as a function of elapsed time for the 
10 ensembles at shell 8.
In the models without a black hole, large $F$'s require both 
$\gamma\approx 2$ and elapsed times in excess of $\sim100$ 
orbital periods.
These conditions are satisfied by many low-luminosity 
ellipticals, which tend to have both steep cusps (\cite{geb96}) and 
dynamical ages in excess of $10^2$ periods (equation \ref{dynam}).
However as $\gamma$ drops, Figure \ref{fig_stat2}a suggests that
the time required to produce a signficant number of 
strongly-evolved orbits quickly exceeds a Hubble time.
For instance, at $\gamma=1$, a time in excess of $10^3$ 
orbital periods is indicated, which is much longer than the 
typical lifetimes of the (bright) ellipticals that have 
weak density cusps.
These results confirm earlier suggestions (\cite{trm96};
\cite{mev96}) that the importance of stochastic diffusion in 
elliptical galaxies without central black holes should fall off 
rapidly with increasing luminosity.
On the other hand, the assumption made in a number of recent 
self-consistency studies (\cite{sch93}; \cite{mef96}; \cite{mer97}) 
of a high degree of chaotic mixing in triaxial galaxies with steep 
cusps, is verified by Figure \ref{fig_stat2}a.

In triaxial galaxies containing a central black hole, 
Figure \ref{fig_stat2}b suggests that $F$ would exceed $\sim 20\%$ 
after 100 orbital periods for any $M_h/M_g\gap 0.005$.
This is roughly the black-hole mass fraction that is believed to be 
characteristic of early-type galaxies of all luminosities (\cite{kor95}; 
\cite{mag98}); among galaxies with well-determined black hole 
masses, galaxies as faint as M32 and as bright as M87 have 
$M_h/M_g\approx0.5\%$ (\cite{vdm97}; \cite{mac97}).
As noted above, even modest black holes, $M_h/M_g\gap0.003$, are
as effective as the steepest density cusps at producing stochastic 
diffusion.
We would therefore expect the degree of dynamical 
evolution in real triaxial galaxies to be determined 
more by the mass of the central black hole than by 
the slope of the central density cusp.
In the galaxies with the largest observed values of $M_h/M_g$,
$M_h/M_g\approx 0.02$, Figure \ref{fig_stat2}b suggests very short 
timescales for evolution, less than 50 orbital periods.
This is as expected given the globally-stochastic
nature of the phase space when $M_h/M_g$ is so large (\S4).

\subsection{Comparison with $N$-Body Evolution}

Having estimated the degree of orbital evolution to be 
expected in triaxial models with a range of structural parameters 
and ages, we would like to go one step further and ask what the 
consequences of this diffusion would be for evolution of a galaxy's shape.
This is clearly a difficult question which can only be properly 
addressed through a combination of self-consistency and $N$-body 
studies.
We will nevertheless attempt to calibrate the expected 
evolution rate against the $N$-body integrations 
of Merritt \& Quinlan (1998), who followed the evolution of a 
triaxial galaxy in which central black holes of various masses 
were grown.
Their initial model was strongly triaxial and 
had $c/a\approx 0.6$, similar to the models considered here.
Merritt \& Quinlan (1998) observed a nearly complete evolution to 
axisymmetry over $\sim 40$ half-mass orbital periods (defined as 
the period of the circular orbit at the half-mass radius in the 
spherically symmetrized model) when the final black hole mass 
was $1\%$ of the galaxy mass.
When $M_h$ was reduced to $\sim0.3\%$, the evolution toward 
axisymmetry (a state that was not quite reached at the end of the integration) 
proceeded at a rate 4-5 times slower.
When $M_h$ was increased to $\sim3\%$, axisymmetry was reached in 
little more than a crossing time.

We can draw some immediate conclusions about the approximate dependence 
of evolution rates on luminosity from this $N$-body work.
Assume, as argued by Magorrian et al. (1998), that the average ratio 
of black hole mass to bulge mass is roughly $0.5\%$, independent 
of bulge luminosity.
Merritt \& Quinlan's results suggest that a triaxial galaxy 
containing a black hole of this mass would evolve strongly in shape
in $\sim100$ periods of the half-mass circular orbit.
According to equation (\ref{dynam}), the half-mass orbital period in 
an elliptical galaxy equals $1\%$ of a galaxy lifetime (of 
$5\times 10^9$ years, say) when $M_B\approx -19$.
Hence we would expect to see strong evolution in the shapes of 
triaxial galaxies fainter than $M_B\approx -19$.

\begin{figure}
\vspace{7cm}
\includegraphics{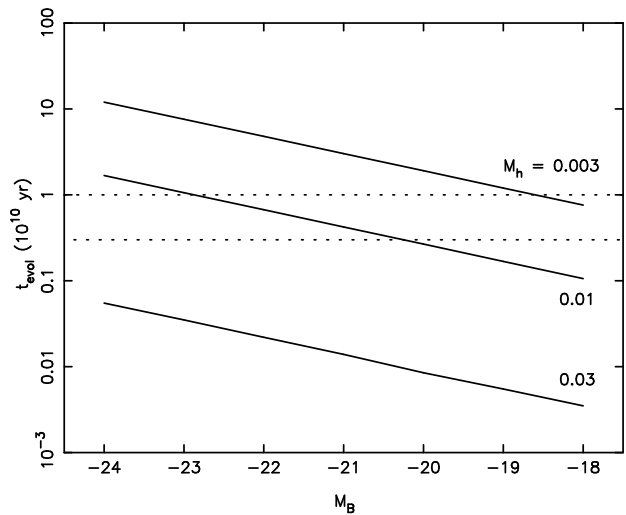}
\caption[]{
Estimated time for strong evolution in the shape of a triaxial galaxy 
as a function of absolute blue magnitude, for various black hole masses.
Dashed lines bracket the expected lifetimes of elliptical galaxies.
}
\label{fig_evol}
\end{figure}

This result is presented graphically in Figure \ref{fig_evol}.
The ordinate in that figure is the evolution time as 
derived by Merritt \& Quinlan (1998) for black 
hole masses of 0.003, 0.01 and $0.03 M_g$; evolution times have 
been converted to years using the scaling relation (\ref{dynam}).
The dashed lines bracket the range of reasonable galaxy lifetimes,
from 3 to 10 billion years.
As just argued, the critical luminosity at which $t_{evol}$ 
equals the expected lifetime of a galaxy is $M_B\approx -19$ for 
fractional black hole masses of $0.5\%$.
However Figure \ref{fig_evol} suggests a rather steep dependence
of this luminosity on $M_h/M_g$, a result of the strong 
dependence found by Merritt \& Quinlan of galaxy evolution rates 
on black hole masses.
Since there is considerable variation in $M_h/M_g$ in real galaxies 
(\cite{kor95}), we would expect that any black-hole-induced correlation 
of galaxy shape with luminosity should exhibit much scatter.
As we discuss below, both the shapes of elliptical galaxies, as 
well as some properties that are believed to correlate with 
shapes, do in fact appear to undergo a change (in the correct sense)
at $M_B\approx -19$ to $-20$.

We would like to take this argument one further step 
and estimate the evolution time 
as a function of the full range of parameters $(\gamma, M_h/M_g)$ 
that were considered in the orbital studies presented above.
We will assume simply that significant evolution occurs when 
some fixed fraction $F_{crit}$ of the stochastic orbits have evolved 
strongly in the sense defined in $\S 5$.
Based on Merritt \& Quinlan's results for $M_h=0.01$, and
scaling their time units to the units of Figure \ref{fig_stat2}
(the scaling factor is 2.3), we find $F_{crit}\approx 0.28$.
If we assume that the same value of $F_{crit}$ can be used to 
predict the evolution toward axisymmetry in models with 
different black hole masses -- probably only a crude approximation -- then 
Figure \ref{fig_stat2} implies an evolution time of $\sim 250T_{1/2}$ for 
$M_h=0.003$, in reasonable agreement with the rate of evolution found 
by Merritt \& Quinlan (1998).
Given this modest success, we feel justified in drawing a dashed 
line in Figure \ref{fig_evol} at $F=0.28$ and inferring 
evolution times for models with $M_h/M_g = 0.001$ and $0.0003$, 
lower than the values considered by Merritt \& Quinlan.
We find in both cases that the predicted evolution times are 
longer than $10^3$ orbital periods.
Only the faintest and densest ellipticals have crossing times 
short enough (equation \ref{dynam}) for such gradual evolution 
to be interesting.
However, such galaxies tend to have steep central density 
cusps (\cite{geb96}), and Figure \ref{fig_stat2}a suggest that
steep cusps would themselves induce evolution at a higher rate.
We conclude that black holes with fractional masses below $\sim 
10^{-3}$ are not likely to be important in producing global evolution 
of their host galaxies.

If we assume that values of $F\approx 0.28$ imply strong evolution 
toward axisymmetry in our black-hole-free models as well, 
Figure \ref{fig_stat2}a implies evolution 
times of $\sim 10^2$ half-mass orbital periods for galaxies with the 
steepest cusps, $\gamma\approx 2$, increasing rapidly for smaller 
$\gamma$.
We conclude that -- in the absence of a central black hole --
only the steepest cusps are capable of inducing significant evolution 
in the shapes of triaxial galaxies over their lifetimes.

\section{Discussion}

Much of the pioneering work on elliptical galaxy dynamics
was carried out in the context of integrable potentials
(\cite{kuz56}, 1973; \cite{del85}; \cite{dez85}). 
More recent work (\cite{sch93}; \cite{mef96}; \cite{mer97})
has emphasized the importance of stochasticity in triaxial 
models with density profiles that mimic those of real elliptical 
galaxies.
The latter studies have typically relied on diagnostics like 
surfaces of section or Liapunov exponents to distinguish regular from 
stochastic orbits.
The frequency mapping technique, first implemented in the context 
of galaxy dynamics by Papaphilippou \& Laskar (1996, 1998), 
represents a significant advance since it allows one to quickly and
accurately calculate the rate at which chaos induces changes in the 
action-angle variables that characterize an orbit.
By contrast, Liapunov exponents reveal only the mean rate 
of divergence in the immediate vicinity of the orbit,
a quantity which may be large even if an orbit is 
confined to a narrow phase-space region over long periods of time.
Frequency mapping also allows the structure of phase space to be 
represented in terms of the quantities that are most important
for the dynamics, the ratios between the fundamental frequencies.

The phase space of triaxial stellar systems is enormously 
complex and our study has revealed only a small part of that 
complexity.
There are many obvious areas for further study.
Figure rotation is a common feature of $N$-body galaxies
and there is some evidence that triaxiality is often associated with 
rapidly-rotating systems like galactic bulges (\cite{kor82}).
While there exists a large body of work on the orbital dynamics of 
rotating barred galaxies (as reviewed by \cite{cog89} and 
\cite{sea93}), much less is known about orbits in rotating triaxial 
systems with steep central density profiles similar to those of galactic bulges.
Another fruitful area for investigation is the character of 
stochastic diffusion.
We argued here that diffusion could be modelled as a random walk 
in frequency space, but this approximation is undoubtedly very 
crude.
One expects to see qualitatively different types of diffusion 
depending on the local structure of phase space (\cite{las93}).
At one extreme, when stochastic layers from many
resonances overlap, the diffusion should be rapid and extensive.
At the other extreme, Arnold diffusion along a single 
resonance line is expected to be extremely slow.

As important as these studies will be for understanding 
the detailed dynamics of triaxial stellar systems, we believe 
that it is already possible to draw two, reasonably secure conclusions 
about real elliptical galaxies.

First, the timescale for stochastic diffusion in triaxial stellar 
systems is often comparable to or shorter than galaxy lifetimes.
Galaxy crossing times vary in a fairly systematic way 
with luminosity (equation \ref{dynam}); to the extent that 
elliptical galaxies of all luminosities begin their lives with 
triaxial shapes, the typical degree of dynamical evolution 
should therefore be greatest in faint ellipticals which have the
shortest crossing times.
One consequence should be a preference for 
more nearly axisymmetric shapes among fainter ellipticals.
We argued that, in galaxies containing a ``typical'' 
black hole with $M_h/M_g\approx 0.5\%$, the dynamical evolution 
time is roughly equal to a galaxy lifetime 
when $M_B\approx -19$.
There is in fact some evidence that the shapes of elliptical 
galaxies undergo a systematic change at about this luminosity 
and that bright ellipticals are moderately triaxial as a class
(\cite{fra91}; \cite{trm96}).
The steep dependence of radio luminosity on absolute magnitude
has also been taken as evidence that bright ellipticals are more
triaxial than faint ellipticals (\cite{aur77}; \cite{bic97}).

Second, we found that the structure of phase space in triaxial 
stellar systems undergoes a qualitative change as the degree of 
central concentration is increased.
When the mass of a central singularity exceeds $\sim 2\%$ the mass 
of the galaxy, a transition to global stochasticity occurs in the 
phase space of boxlike orbits.
One would expect to observe a very rapid evolution in the shape 
of a triaxial galaxy when this critical mass is exceeded,
since in the globally-stochastic regime, boxlike orbits 
lose their distinguishability in just a few crossing times.
Merritt \& Quinlan (1998) have in fact verified that an initially 
triaxial galaxy evolves toward an axisymmetric state in little more than 
a crossing time when $M_h/M_g$ exceeds $\sim 2.5\%$.
Remarkably, real galaxies also seem to know about this critical mass 
ratio: the distribution of $M_h/M_g$ (with $M_g$ defined as the mass of 
the bulge in spiral galaxies) appears to fall off sharply above $\sim 
2\%$ (\cite{mag98}), and the two galaxies with the largest ratios 
of black hole mass to bulge mass, NGC 3115 and NGC 4342, 
both have $M_h/M_g\sim 0.025$ (\cite{kor96}; \cite{vdb97}).
Merritt \& Quinlan (1998) argue that this agreement is more than a 
coincidence.
The fueling of massive black holes in quasars and AGN requires 
matter to be funneled into the nucleus from large distances, and 
a number of lines of evidence suggest that gravitational torques 
from non-axisymmetric perturbations are a necessary ingredient 
(\cite{shb90}).
A sudden transition to axisymmetry in the stellar distribution
would therefore be expected to limit the mass of a central black hole.

We are tempted to incorporate these ideas into a more complete 
picture of elliptical galaxy evolution, as follows.
We suppose that elliptical galaxies begin their lives as gas-rich 
disks.
The bulges of these disk galaxies contain black holes, which form 
from gas that is channeled into the nucleus by gravitational torques 
from the (triaxial) bulge and/or stellar bar.
This growth is halted (presumably at the end of the quasar epoch, 
at $z\approx 2-3$) when the black holes have accreted $\sim 2\%$ 
the mass of their host bulges, at which point the infall of 
matter comes to a halt as the bulges acquire axisymmetric shapes.
Subsequent mergers between disk galaxies convert many of them into 
ellipticals; the black holes also merge, coming to rest in the 
nuclei of the merged galaxies.
Mergers have two important consequences for the subsequent dynamics.
First, they convert gas into stars and disks into spheroids, thus reducing the 
average value of $M_h/M_g$ below the critical value that 
generates global stochasticity in the bulge -- perhaps to $\sim 0.5\%$, the 
average value observed in early-type galaxies in the nearby universe.
Second, mergers create triaxial systems from initially 
axisymmetric ones.
Taken together, these two facts imply different end states for 
bright and faint ellipticals.
Bright ellipticals, with their low central densities and long 
crossing times, would tend to retain their (merger-induced)
triaxial shapes.
Faint ellipticals, with their high central densities and short 
crossing times, would remain close to axisymmetric.
The result, at the present epoch, would be an apparent 
dichotomy, with bright, triaxial ellipticals at one extreme and faint, 
axisymmetric ellipticals at the other.

The existence of two families of elliptical galaxies with distinct 
morphologies has in fact been emphasized by Kormendy \& Bender (1996), 
who argued that elliptical galaxies should be divided into two groups, 
disky and boxy, based on the deviations of their isophotes from ellipses.
We note that strong boxiness is a generic feature of $N$-body galaxies,
whether formed via collapse (\cite{udr93}), dynamical instabilities 
(\cite{rah91}), accretion (\cite{qug86}), mergers (\cite{her90}),  
tidal torquing (\cite{mav85}), etc.
The ubiquity of boxiness in $N$-body galaxies is a consequence of the fact 
that regular orbits, both tubes and boxes, have dimpled 
shapes when seen in projection.
A galaxy constructed from such orbits is likely to be boxy too
unless the distribution of orbital turning points is chosen to be 
sufficiently smooth (\cite{bip85}).
The interesting question is: How do some elliptical galaxies 
{\it avoid} being very boxy?
One plausible answer, suggested by Figure \ref{fig_orbits}, is 
that stochastic diffusion eliminates sharp features in the turning-point
distribution of the boxlike orbits.
$N$-body simulations (\cite{frb93}; \cite{meq98}) provide some 
support for this idea: the growth of a central mass concentration
can convert a boxy, triaxial system into an
axisymmetric one with accurately elliptical isophotes.
We therefore propose that boxiness is an 
indication that the phase-space distribution has not been strongly 
influenced by stochastic diffusion.
In support of this view, we note Kormendy \& Bender's (1996) 
observation that boxiness correlates well with 
kinematical measures of triaxiality (\cite{cal94}), consistent with 
our expectation that triaxiality itself
can only be maintained in systems that are dynamically unevolved.

Boxy systems constructed primarily from tube orbits -- an example 
is shown in Figure 12 of Sellwood \& Merritt (1994) -- would not be 
strongly affected by stochastic diffusion.
This fact might explain the persistence of strong boxiness in the 
bulges of many disk galaxies (\cite{sha87}).

It is often argued (e.g. \cite{kor90}; \cite{fab97}) that many of 
the systematic 
differences between bright and faint ellipticals are due to the greater 
importance of gaseous dissipation in the formation of the latter.
We do not disagree with this view but point out that one of 
the principal effects of dissipation is to accelerate
{\it stellar dynamical} evolution of galaxies by increasing their 
mean density and central concentration.
For instance, $N$-body simulations of triaxial or barred galaxies 
containing a dissipative component often show a sudden change in the 
galaxy's shape, toward axisymmetry, after a small fraction of the 
gas has accumulated in the center (\cite{udr93}; 
\cite{dub93}; \cite{frb93}; \cite{bah96}).
While sometimes attributed loosely to ``dissipation,'' this rapid 
evolution in the stellar distribution can only be due to a 
change in the character of the orbits resulting from the
increased central force.
In other words, it is an indication that many of the stellar orbits 
have become chaotic.
Of course, dissipation has other effects as well, most importantly an 
enhancement of rotational support.
Our point here is that a trend of increasing dissipation with 
decreasing luminosity is not inconsistent with the view that faint 
ellipticals are dynamically more evolved than bright ellipticals
as well.

\bigskip

This work was supported by NSF grants AST 93-18617 and AST 96-17088
and by NASA grant NAG 5-2803.  
I. J\o rgensen kindly provided the data on which Figure \ref{fig_inger} 
is based.  
We thank J. Laskar, R. Nityananda, Y. Papaphilippou, D. Richstone,
J. Sellwood and S. Tremaine for useful discussions.  
H. Kandrup critically reviewed the initial manuscript and his comments led
to substantial improvements in the final version.

\end{document}